\documentclass[12pt,a4paper]{article}
\usepackage[margin=1in]{geometry}
\usepackage[utf8]{inputenc}
\usepackage{natbib}
\usepackage{hyperref}
\usepackage{verbatim}
\usepackage{amsmath}
\usepackage{amsfonts}
\usepackage{amssymb}
\usepackage{amsthm}
\usepackage{mathtools}
\usepackage{bm}
\usepackage{xparse}
\usepackage{graphicx}
\usepackage{xcolor}
\usepackage{textcase}
\usepackage{url}
\usepackage{lipsum}
\usepackage{appendix}
\usepackage{epstopdf}
\usepackage{setspace}
\usepackage{centernot}
\usepackage{braket}

\usepackage{titlesec}

\usepackage[pagewise]{lineno}

\titleformat{\section}
  {\normalfont\scshape \filcenter \large}{\thesection}{1em}{}

\titleformat{\subsection}
  {\normalfont\scshape \filcenter}{\thesection}{1em}{}

\usepackage{indentfirst}

\DeclareMathOperator{\Tr}{Tr}

\usepackage{amsthm}

\usepackage{authblk}

\usepackage{dsfont}

\begin{document} 

\title{Notes on a future quantum event-ontology}

\author[1, 2]{Sebastian Horvat}
\affil[1]{University of Vienna, Department of Philosophy, Universitätsstraße 7, 1010 Vienna, Austria}
\affil[2]{University of Vienna, Faculty of Physics, Boltzmanngasse 5, 1090 Vienna, Austria}
\date{}                     
\setcounter{Maxaffil}{0}
\renewcommand\Affilfont{\itshape\small}


\maketitle

\begin{abstract}
This essay is a two-step reflection on the question `Which events (can be said to) occur in quantum phenomena?' The first step regiments the ontological category of \textit{statistical phenomena} and studies the adequacy of \textit{probabilistic event models} as descriptions thereof. Guided by the conviction that quantum phenomena are to be circumscribed within this same ontological category, the second step highlights the peculiarities of probabilistic event models of some non-relativistic quantum phenomena, and thereby of what appear to be some plausible answers to our initial question. The reflection ends in an aporetic state, as it is by now usual in encounters between ontology and the quantum.
\end{abstract}

\bigskip

\bigskip

The vast amalgam of cognitive-material activities that comprise the empirical sciences of the last few centuries bears a distinctively \textit{statistical} character. This is so not only for those sciences that deal with human society and history - which is so natural for us nowadays to perceive as messy, irregular, contingent - but also for those that thematize the allegedly most fundamental and ahuman aspects of the world, exemplified by classical statistical mechanics and quantum theory. The \textit{non}-statistical science that was previously advocated by the likes of Galilei, Descartes and Newton perceived individual objects, events, phenomena and their relations as instantiations of categorically assertable orders: the movement of \textit{this} particular rock falling from the tower of Pisa is to be viewed as \textit{an} instance or sign of an ideal mathematically presentable order that holds transhistorically for such-and-such bodies moving in such-and-such conditions. The particulars surrounding the early modern scientists were accordingly imperfect images of an ideal mathematical order, arguably echoing the ancient doctrine of Forms propounded by Plato and the mathematico-cosmological speculations of the Pythagoreans. The statistical sciences that later emerged posed a different relationship between the particular and the ideal, in that it is not individual objects, events, phenomena or relations that signify the ideal, but intelligently selected pluralities thereof. Schematically stated, the Aristotelian scientific unit `Each X is Y' is replaced by the less assertive `Some X are Y', now mathematizable in terms of the newly developed theories of probability. The world that confronts statistical science is thus still rational, albeit in a different and an arguably weaker sense.

Among the objects of study of the statistical sciences we find spatially and temporally extended phenomena - henceforth, \textit{statistical phenomena} - wherein the units that exhibit statistical regularities are spatiotemporally localized events. For example, a simple coin toss can be dissected into the initial movement of the hand that is to execute the toss, the orientation of the coin in mid-air and the final landing of the coin, whereby the latter two exhibit statistical regularities that can be modelled by appropriately chosen probability distributions. A scientifically more interesting class of examples is given by \textit{quantum phenomena}, that is, those statistical phenomena that are usually accounted for with the use of quantum-mechanical models. The by-now paradigmatic electronic double-slit experiment features statistical regularities of the detection of electrons at various locations on a screen, which coincide with the empirical predictions of an appropriate use of quantum theory. There is however an important difference that distinguishes current theoretical accounts of quantum phenomena from theoretical accounts of other statistical phenomena, e.g. those studied by classical statistical mechanics. The difference lies in the lack of clarity on which events take place in quantum phenomena, besides those few events that quantum mechanical models are already concerned with. Whereas there is consensus that in each execution of the said double-slit experiment there are events that correspond to the detection or non-detection of the electron at some location on the screen, there is no consensus on which events, if any, take place in the void that is allegedly traversed by the electron before impinging on the screen. In other words, there is no agreement on which ``event-ontology'' is adequate for quantum phenomena, paralleling the lack of consensus on which, if any, ``object-property-ontology'' would be so adequate.\footnote{A cursory critical evaluation of what some of the currently dominant proposals imply for an event-ontology of quantum phenomena will be given in Section IV.II. For a more general introduction to disagreements concerning the interaction between ontology and the quantum, see Lewis (2016).}

This essay is a reflection on the question: `Which events occur in quantum phenomena?' This problem may be brushed off too easily as a remnant of an old traditionalist metaphysical striving that is to be eliminated or dissolved by more liberating pragmatist-instrumentalist ways of thinking. I do however regard it as a genuine problem, not because of some conservative holding on to the metaphysical tradition, but out of the basic hermeneutic need of understanding quantum phenomena in terms in which we seemingly unproblematically understand other statistical phenomena. Coin tosses and double-slit experiments after all do occur in the same world, and deserve an attempt at being parallelly treated as examples of one and the same general ontological category. The aim is thus physiological: it is to digest the problematic in terms of the unproblematic, the quantum portion of the scientific image in terms of the statistical phenomena portrayed in the manifest image. Besides the satisfaction of this hermeneutic need, a resolution to our problem might - and here emphasis on the `might'! - also ease the satisfaction of the desire for an integration of quantum theory and relativity theory. An investigation into our question may indeed ideally lead to quantum-mechanically inspired principles that concern events and their statistical relations, which might be manageably integrable with principles underlying relativity theory, the latter being itself a theory concerned with \textit{spatiotemporal} relations among events. Whether this promisory note has any merit to it can however only be evaluated after our question is systematically addressed.

The hermeneutic digestion that is called for above demands an exercise in ontology-metaphysics, and as such, it risks no less than other such exercises in ending up producing lifeless schemes whose internal logics are cut off from the rest of science, let alone from the rest of our lives. The ontologist ought to be weary of proclaiming the existence or occurrence of a certain X without critically relating this X as clearly and closely as possible to the familiar. She should be weary, that is, of forging a pathological ``realism'' that forgets that its aim is not just to look for a formalistic ``truth'' in the abstract, but for an intelligible truth. A danger is otherwise faced of lapsing into technically sophisticated but ultimately philosophically non-rigorous schemes that metaphysics is by its very nature prone to and that have been recurrently repelling generations of philosophers and scientists since its ancient beginnings. 

In an attempt of abiding by this roughly sketched ideal typus, we will first discuss statistical phenomena in general, keeping in mind ordinary examples such as coin tosses, and analyzing the adequacy of models that can be used to describe such phenomena. Only then will we apply the general discussion to quantum phenomena, attempting to treat them as ordinarily as possible, suppressing our quantum-mechanically trained prejudices on how to interpret them. The reader is to be warned that the reflection will not lead here to any definite verdict on which, if any, event-ontology is adequate of the vast plurality of phenomena that are studied with the aid of quantum-mechanical techniques. The essay will instead prepare the ground for a possible such future verdict (sections I-III) and bring to the fore some peculiarities that concern a possible event-ontology of a subclass of quantum phenomena (section IV). The hope is nevertheless that what follows points at least preliminarily at the right direction.

\section*{I. Statistical Phenomena}

Ontology, no less than any other science, cannot encounter its objects of study without some preliminary sweat - without some aiming, smoothening, removing, adjusting, (de)forming. Accordingly, the family of phenomena that we will here aim to encounter and subsume under a unified smoothened ontological category comprises a heterogeneity of objects that share some core commonalities, but also exhibit many differences, and are therefore studied by different special sciences. In what follows, we will take a look at some examplars of our yet-to-be rectified category of \textit{statistical phenomena} and we will highlight some of the important features that they all appear to share.

Statistical phenomena (SP), besides nowadays being omnipresent in our daily concerns, are widely studied in the natural and social sciences, examples being: the toss of a coin and its subsequent landing on a table, the injection of proton beams into a particle-collider and the subsequent detection of a high energy photon under some angle, a medical test in which a drug is injected into a middle-aged patient and some health-parameter is registered, an action performed by the Austrian National Bank and the consequent change in inflation. All of these examples exhibit regularities at the level of collections, repetitions or populations, the members of which all can be said to be instances of the same phenomenon. A fair coin does not always land heads nor does it always land tails, but manifests a regularity in the proportion of heads and tails in relatively large collections of coin tosses, whereby the various tosses need to be similar to each other in various respects, such as in the constitution of the tossed coins, the methods of tossing, the surrounding gravitational fields, and so on. This already points to an important feature of SPs, that is, that all of them need to admit of at least some \textit{non-statistical} determination. This is because any SP can empirically manifest its regularity only at the level of a population, which presupposes that a plurality of different particular sequences of events be identifiable as members of that population. This identification, if it is not to rely on tacit knowledge of the sort that may tentatively be attributed to a neural network, presupposes the possibility of an explicitly articulated description that is ideally either verified or not by any particular sequence of events. In short: different concrete sequences of events can be identified as instances of \textit{the same} phenomenon only if they share something articulable in common. A ``purely statistical'' phenomenon - one whose instances need not share anything articulable in common - cannot become an object of empirical scientific investigation. Usurping Kantian vocabulary, this condition may be cast as ``transcendental'' - as a condition of the possibility of a SP being presentable within our scientific experience. Let us summarize it in form of a principle. 

\bigskip

\textbf{Principle 1.} Every SP necessarily admits of a non-statistical linguistic-mathematical description, that ideally either applies or does not apply to any particular sequence of events, and thereby categorizes it as being or not being an instance of that SP.

\bigskip

We will henceforth refer to such descriptions as \textit{identifiers}. It is important to note the word `ideally' appearing in the above principle. No actual identifier is a \textit{perfect} identifier, that is, one that could fully unambiguously mark a particular here-and-now as being or not being an instance of the SP that it is aiming to identify. The identifier of a regular coin toss may specify the sort of coin that is to be used, the type of machine that is to execute the toss, the type of table on which the coin is to land, the strength of the gravitational field and the pressure of the air that are to surround the coin; and yet, despite all these determinations, it might still fail to absolutely unambiguously adjudicate whether \textit{this} toss of \textit{this} coin in \textit{these} circumstances counts as an instance of a fair coin toss or not.\footnote{Here is an example. A precisification of the fairness of a coin might for instance appeal to the unambiguously definable symmetry of its mathematically representable mass distribution $\rho(\vec{x})$. The ambiguity however derives from the fact that \textit{this} material coin that is currently in my wallet does \textit{not} have a fully unambiguously attributable mathematically representable mass distribution. Where is the infinitely thin boundary that separates my coin from its surroundings? The ambiguity of the answer to this question leaks into the assessment of which $\rho(\vec{x})$ best describes the coin.} This is a mere instance of the sort of gap that necessarily persists in any attempt of relating a description - be it mathematical or not - to a concrete material here-and-now, such as in relating a description of what-a-bottle-is to a particular material object. This indefiniteness of the particular, juxtaposed to the definiteness of the abstract, can nevertheless be for the most part ignored in ordinary life and in scientific practice: barring pathological cases, agreement is usually reachable on whether \textit{this} object is to be categorized as a bottle and whether \textit{that} series of events is to be regarded as a toss of a fair coin.\footnote{This usual reachability of agreement concerning the applicability of a certain description or concept to a particular possibly-here-and-now is in no way accidental - as if, history and prehistory having been different, we could have ended up wielding concepts whose reference wildly oscillates from wielder to wielder. Concepts, descriptions and identifiers are publicly available tools that are ideally usable by an anonymous \textit{anyone} - especially so in science, the cluster of subjectivity-suppressing practices \textit{par excellence} - and as such, bear functions that are forged socially, and not by an arbitrary intention of an individual will. Note the qualifiers `usual', `for the most part', `barring pathological cases', and so on, which are a reminder that we are here operating at the slippery boundary between the abstract and the concrete, the conceptual and the pre-conceptual, the formal and the material. The possibly-here-and-now is inexhaustible by our conceptual activities, it is so-to-say always a step ahead of our grasp, which is in turn what makes genuine creation and revolution possible. Totally unobstructed creativity is destined to turn into paralysis. Fortunately (?) however, all creations, be that they are propounded by philosophers, scientists, poets or totalitarian leaders, are bound to show cracks, to lose their originary significance, and to give birth to new creations. They are born, they live (in)glorious lives and they die, as all of us do. But they also leave a trace behind, as all of us - hopefully (?) - do.


} 

Note that, besides the constraint of not referring to statistical distributions, the content of SPs' identifiers is left unspecified and open. Identifiers might be phrased linguistically as in `a toss of a fair coin', but may also refer to mathematical objects, as in `a toss of a coin with mass distribution $\rho(\Vec{x})$' or `a measurement of the momentum of an electron in quantum state $\ket{\alpha}$', where $\rho(\Vec{x})$ is a function from $\mathbb{R}^3$ to $\mathbb{R}_+$, and $\ket{\alpha}$ is an element of a complex Hilbert space. Identifiers also cut across the ``observable-unobservable'' divide: they might refer to exotic theoretical posits such as wave functions and gauge potentials, or to mundane things and events, such as loud sounds and falling rocks. An identifier is what it is purely in virtue of its function, which is to enable the identification of a particular sequence of events as an instance or non-instance of a SP, a function that is implementable with potentially hybrid linguistic-mathematical theoretical-mundane descriptions.

\section*{II. Probabilistic Event Models}

Besides what has just been said about identifiers, there is another important property that is shared by all SPs. In so far as they are spatially and temporally extended, all SPs appear to be analyzable into events taking place in relatively localized spatial and temporal regions. This is obvious for SPs studied by physics - especially so for those where the spatiotemporal determinations of the events in question are explicitly thematized in their canonical physical models, as it is the case for special-relativistic phenomena - but it also holds for all other SPs, including those that are studied by the social sciences. Take for instance the example of the Austrian National Bank's action and the consequent change in inflation. Whereas the two events under consideration are more abstract and complex than a coin toss, they are nonetheless related to events that take place in Austria within a few months or years, rather than, say, in Athens within a span of a couple of centuries. A SP can accordingly roughly be characterized by (i) a list of spatiotemporal regions, (ii) the events that can occur in these regions, and (iii) their statistical distribution. We will assume here that the spatiotemporal regions and the relations between them can be adequately represented by the flat Minkowski spacetime $(\mathbb{R}^4, \eta)$, where elements of $\mathbb{R}^4$ represent infinitesimal regions, and $\eta$ is the Minkowski metric tensor that quantifies spatiotemporal relations between different regions.\footnote{Here is some notation that will occasionally be used later. The causal past of region $X \subset \mathbb{R}^4$ will be denoted by $\mathcal{C}^{(-)}_X \equiv \left\{y \in \mathbb{R}^4| \exists x \in X: y \prec x \right\}$, where $y \prec x$ holds if there is a future-oriented time-like or null-like curve from $y$ to $x$, and $y \neq x$. Analogously, the causal future of a region $X$ is $\mathcal{C}^{(+)}_X \equiv \left\{y \in \mathbb{R}^4| \exists x \in X: x \prec y \right\}$. Finally, the set of points causally unrelated to $X$ is $\mathcal{C}^{(0)}_X\equiv \mathbb{R}^4 \setminus \left(\mathcal{C}^{(+)}_X \cup \mathcal{C}^{(-)}_X \cup X \right)$.} We will thereby restrict ourselves to a treatment of only those SPs in which gravitational effects, if present at all, can be captured without general relativistic techniques. 
This being said, let us introduce a general class of mathematical structures and models that are intended to be usable as descriptions of any (non-general-relativistic) SP. 

\bigskip

\textbf{Definition 1.} (Probabilistic event structure) Let $\left\{\Omega_x \right\}_{x \in \mathbb{R}^4}$ be a family of sets and let $\mathcal{X}$ be the family of all finite subsets of $\mathbb{R}^4$. For each $X \in \mathcal{X}$, let $\mathcal{F}_X$ be the $\sigma$-algebra on $\Omega_X$, where $ \Omega_X = \prod\limits_{x \in X}^{} \Omega_x$. A \textit{probabilistic event structure} is a family of probability measures $\left\{(\Omega_X,\mathcal{F}_X, \mu_X) \right\}_{X\in \mathcal{X}}$, such that for all $X' \subset X$, $\mu_{X'}$ is the corresponding marginal measure of $\mu_{X}$.

\bigskip

A probabilistic event structure (PE-structure) is thus a family of mutually consistent probability measures indexed by finite subsets $X$ of $\mathbb{R}^4$, whose corresponding sample spaces are products of those sample spaces that are associated to points $x \in X$. In order for these structures to be used as tentative descriptions of SPs, the mathematical objects that figure therein are to be suitably related to something non-mathematical. We will accordingly introduce \textit{probabilistic event models} as interpreted PE-structures. 

\bigskip

\textbf{Definition 2.} (Probabilistic event model) An interpreted PE-structure $\left\{(\Omega_X,\mathcal{F}_X, \mu_X) \right\}_{X\in \mathcal{X}}$ is a \textit{probabilistic event model} if the elements of $\Omega_X$ are taken to represent ordered lists of events that can occur in spatiotemporal regions $X$, and each $\mu_X$ is taken to represent the probability distribution of events $\Omega_X$. 

\bigskip

Probabilistic event models (PE-models) assign probability distributions to events that occur in any finite collection of infinitesimal spatiotemporal regions and thus appear as good candidates for being used in the describing of SPs. PE-structures - and consequently, PE-models - prove to be rather compactly presentable, owing to a straightforward application of \textit{Kolmogorov's extension theorem}, as summarized in the following proposition.\footnote{For a presentation and proof of the theorem, see Tao (2011; 196-198).}

\bigskip

\textbf{Proposition} (Application of Kolmogorov's extension theorem) Let $\left\{(\Omega_X,\mathcal{F}_X, \mu_X) \right\}_{X\in \mathcal{X}}$ be a PE-structure. Let $ \Omega = \prod\limits_{x \in \mathbb{R}^4}^{} \Omega_x$ and $\mathcal{F}$ be the product $\sigma$-algebra of spaces $\left\{(\Omega_x,\mathcal{F}_x) \right\}_{x\in R^4}$. The PE-structure then determines a unique probability measure $\mu$ on $\left(\Omega, \mathcal{F}\right)$, whose marginal measures on $\Omega_X$ coincide with $\mu_X$, for all $X \in \mathcal{X}$.

\bigskip

A PE-model can thus be compactly represented by a unique measure $\mu$, that encodes the ``empirically accessible'' distributions $\mu_X$, for finite subsets $X \subset \mathbb{R}^4$. When referring to a PE-model, we will thus henceforth refer only to its induced distribution $\mu$, while retaining $\mu_X$ for its marginal distributions on $X$. 
Now, note that PE-models assign distributions to each and every finite collection of spatiotemporal regions, amounting to a positing of more structure than is needed to model the SPs that are usually of interest in the natural and social sciences. For example, in modelling a particle collision experiment, physicists are interested only in a handful of spatiotemporal regions, such as the one enveloping the emission and detection of the collided particles, but not e.g. those that contain events that take place on the Moon two centuries after the experiment. More precisely, to any SP we can associate a particular collection of \textit{regions of relevance} $X \subset \mathbb{R}^4$, such that only the random events that take place therein have a determinate statistical distribution, while the distribution of the remaining events is undetermined or arbitrary. Consequently, any two PE-models that agree in their marginal distributions on $X$ are at least \textit{prima facie} equally adequate as models of that SP. All this spuriousness and consequent non-uniqueness is nevertheless retained for the sake of generality, in that we are here aiming to define a class of models, the elements of which can in principle be used to describe any (non-general-relativistic) SP\footnote{Spurious structure is hereby retained for the sake of theoretical elegance and unification. This is hardly new neither in ontology nor in physics: wholeness and unity of the abstract are made possible only by widening the distance from the particular and from its complexity. The elegance of Maxwell's equations implies the intricacy of their application to any particular possibly-here-and-now, as every physics student and electrical engineer can testify. It is the artistry of the theoretician to find a healthy balance between the two, a balance between the self-enclosed and unitary principle, and the manifold of disordered, disunified particulars - or, put even more schematically, between what may be called the order of ``pure form'' and the disorder of ``pure matter''.} - with another potential perk of this spuriousness being that a general-relativistic modification of PE-models might (!) come of use in cosmology, whereby the history of ``the cosmos'' may tentatively be regarded as a particular instantiation of some SP whose region of relevance $X$ is taken to cover ``the whole of space and time''.\footnote{The scare quotes are a reminder of the problematicity of the limit-concepts of ``the cosmos'' and of ``the whole of space and time''. There seems to be an important difference between, on one hand, the elements of an ordered series of SPs with ever increasing spatiotemporal extensions, and on the other, the supposed limit(s) of this series. An investigation of this difference and of the true sense of this limit are needed for an adequate understanding of the meaning of physical cosmology - at least of a cosmology that aims to study ``the cosmos as a whole''.} Finally, note that the collection of relevant spatiotemporal regions $X$ associated to some SP might often be only vaguely definable, owing to the vagueness of that SP's identifier. In the above example of the particle collision, while it is obvious that the events taking place on the Moon two centuries after the experiment are of no relevance, it is not determined where exactly the region of relevance ends at an infinitesimally minute level - e.g. whether it ends at location $x$ or at location $x+\epsilon$, for some infinitesimal $\epsilon>0$ - this once again being a manifestation of the aforementioned gap between the abstract and the concrete that is for the most part effortlessly and unproblematically bridged by scientific-experimental practice. 

Recall that the content of SPs' identifiers was beforehand left unspecified, as long as they fulfilled their basic function of marking particular sequences of events as instantiations of SPs. However, in all of the examples referred to above - and more broadly, in examples that are of interest in the empirical sciences - identifiers refer to events that take place outside of the regions of relevance. The identifier of a coin toss specifies the conditions that need to obtain prior to the landing of the coin, whereas the identifier of a quantum measurement of a particle consists in a specification of the states of the particle and of the measurement device prior to the outcome of their interaction. Indeed, if identifiers were not restricted in this way and were allowed to refer to events that happen within a SP's region of relevance, then the category of SPs would end up being too broad for our current scope. For instance, the class of valid identifiers would then also include `the toss of a fair coin that lands heads', with the region of relevance including the landing of the coin. A theory of so-identifiable phenomena would not track statistical regularities, but mere regularities of the possibility of co-occurrence of events. Since we are here not interested in the latter, we will assume that those regions that host events referred to by identifiers do not overlap with the regions of relevance. Furthermore, we will here be focused only on SPs that are identifiable by reference to events that take place specifically in the \textit{causal past} of their pertaining region of relevance $X$, i.e. somewhere in $\mathcal{C}^{(-)}_X$. The obtaining of events that verify the identifier of one such SP can thus be thought of as ``probabilistically causing'' the taking place of events in that SP's region of relevance. It should be noted however that it might as well be interesting to study those SPs that are not identifiable without reference to e.g. the causal future of their regions of relevance: think for instance of a convoluted quantum experiment involving a series of measurements wherein a specific outcome of the final measurement is postselected.

\section*{III. Adequacy}

The function of PE-models is to describe or represent SPs by inducing a story about which random events occur therein: in what follows we will draw some principled constraints on PE-models from the basic requirement of them possibly exercising this function. 

Whether a particular PE-model correctly-adequately-satisfactorily describes a given SP evidently depends somehow on the relationship that obtains between what the PE-model asserts about the regions of relevance $X$ and what really happens in regions $X$ in the given SP. Slightly more substantially, a PE-model $\mu$ is an adequate description of SP $Z$ if its posited marginal distribution $\mu_X$ approximately coincides with the statistical distribution of the events localized in $X$ as they actually unravel in phenomenon $Z$.\footnote{There is a sea of considerable complexity lying beneath `the statistical distribution' that the model is supposed to (approximately) coincide with. The transformation of actual material instances of SPs (these coin tosses here and now) into a mathematical form that can potentially be compared to a mathematical model ($p=\frac{1}{2}$) requires plenty of cognitive-material work: measuring, selecting, smoothening, discarding, etc. For more details on this, see some contemporary literature on scientific representation, e.g. Van Fraassen (2008) and Nguyen \& Frigg (2022).} The possibility of ascertaining whether $\mu$ is an adequate description of $Z$ thus presupposes the possibility of an empirical determination of the sought statistical distribution that the model is to be compared to. This in turn presupposes that for each spatiotemporal region $x \in X$, it is possible to verify which event in particular among the ones referred to by $\Omega_x$ is instantiated in any given instance of $Z$. For example, in order to ascertain whether a particular probability distribution correctly captures the statistics as it actually unravels in a series of coin tosses, it is obvious that one needs to be able to ascertain for any particular toss whether it has landed heads or whether it has landed tails. Now, this possibility of verification generally depends on the capabilities of the one who is doing the verifying, e.g. on the person or group of people that are using certain instruments to check which event happened at a given location $x$. Verifiability is a notion that is relative to verifiers - to particular agents and instruments employed to conduct the verification - and its requirement therefore does not amount to a precisely phrasable general constraint on PE-models. This dependence on contingently chosen verifiers is only seemingly remedied by imposing the stronger limit-requirement of ``in-principle-verifiability'', which would demand verifiability by idealized agents equipped with ideal instruments: however, despite its apparent generality (its reference to \textit{any} agent and instrument), it remains for the time being too vague to amount to any clear constraint on PE-models (and is probably bound to remain so).\footnote{The notions of agency and instrumentation do not have precise aculutural-ahistorical-acontextual circumscriptions. X is an instrument/agent if X is taken-to-be-an-instrument/agent within a particular cultural context. Whereas it is not up to us whether an electron or a tree are present in this room, it is partly up to us whether an agent or an instrument are here present. After all, we do appear to live in an age that anticipates a society filled with cultural contexts that, quite perversely, turn human beings into instruments and artificially constructed machines into agents.}

Nevertheless, despite its inexactness, verifiability to the very least implies another condition, which does impose a precise physical constraint on PE-models, in that it replaces talk of agents and instruments by talk of \textit{physical possibility}. Consider a PE-model $\mu$ that is an ascertainably adequate description of a SP $Z$, and that asserts that a random event at location $x$ is distributed according to probability distribution $\mu_x$ on $\Omega_x$. As stated beforehand, that the adequacy of the model is ascertainable implies the existence of a procedure that can be used by an (ideal) agent to verify which event in $\Omega_x$ has taken place at $x$ in any concrete instance of $Z$. Now, a particular execution of this procedure that confirms that some event $\omega^* \in \Omega_x$ has happened at $x$ is an instance of another SP $Z^{(x,\omega^*)}$, whose events in the causal future of $x$ are distributed almost equally as in $Z$, except for the condition that event $\omega^*$ took place at $x$. That is, a model $\mu^{(x,\omega^*)}$ that is adequate of $Z^{(x,\omega^*)}$ asserts that the random events in the future of $x$ are distributed according to $\mu^{(x,\omega^*)}_X(\omega_X)=\mu_{X}(\omega_X|\omega_x=\omega^*)$, for any $X\subset \mathcal{C}^{(+)}_x$. Therefore, the (ideal) possibility of verifying that $\omega^*$ occurred at $x$ presupposes to the very least the physical possibility of phenomenon $Z^{(x,\omega^*)}$.
A similar conclusion follows from the requirement of the conjoined verifiability of a plurality of events, summarized in form of the following principle.

\bigskip

\textbf{Principle 2.} (No absolute indeterminism) Let $\mu$ be an adequate model of some SP $Z$. Then, for any finite $S \subset \mathbb{R}^4$, and $\omega \in \Omega_S$ such that $\mu_S(\omega)\neq 0$, there is a physically possible SP $Z^{(S,\omega)}$ that is adequately modelled by $\mu^{(S,\omega)}$, which satisfies
\begin{equation*}
    \mu^{(S,\omega)}_X(\omega_X )=\mu_{X}(\omega_X|\omega_S=\omega),
\end{equation*}
for all $X \subset \mathcal{C}^{(+)}_S$ and $\omega_X \in \Omega_X$.
 
\bigskip

Stated succinctly, the above principle says that a commitment to something possibly happening in a SP entails a commitment to the possibility of this something definitely happening in another SP. A responsible assertion that $\omega$ can \textit{possibly} take place at $x$ and that the events in $X$ can \textit{possibly} be distributed according to $\mu_{X}(\omega_X|\omega_S=\omega)$ requires the commitment to the possibility of $\omega$ \textit{definitely} taking place at $x$ and the events in $X$ \textit{definitely} being distributed according to $\mu_{X}(\omega_X|\omega_S=\omega)$. In other words, there is \textit{no absolutely indeterministic event} - all chance is contingent on what goes on in the causal past of the event. The chancy nature of the outcome of the coin toss derives from the conditions in which the coin is tossed - there obviously exists a physically possible SP in which the coin simply lays on the table heads up. The same holds for measurement outcomes in quantum-mechanical experiments, despite their seemingly irreducible indeterminism: even though the chancy outcome of a measurement of an electron's momentum seemingly cannot be understood as deriving from a lack of control of certain microscopic degrees of freedom, it is still the case that for any possible measurement outcome, there is a SP wherein this outcome takes place deterministically. No chance is absolute, or conversely, all chance is contingent. As stated earlier, the requirement of this principle is implied by the requirement of verifiability, and this is so essentially in virtue of technological possibility - ``our'' possibility of doing thus and so -
implying physical possibility - the possibility of something being or happening thus and so. Whereas (in-principle-)verifiability appeals to arguably non-physically definable notions of agency and instrumentation, the above principle recurs only to the notion of physical possibility, a few words on which are now due. 

The physical possibility of the realization of a SP is here understood as being an objective physical determination thereof: that particle-interference is physically possible and that superluminal signalling is physically impossible is as independent of us and our theorizing as the fact that the dinosaurs were exterminated by the impact of a meteorite. Even though the description of a statistical phenomenon may take recourse to theoretically posited notions, whether this phenomenon is possible or not is independent of theory. A particular experiment that might have previously been interpreted as enabling the measurement of the gravitational force between two bodies is physically possible regardless of whether there is such a thing as the ``gravitational force''. Conversely, particle-interference was physically possible also before physicists constructed the material and conceptual means that enabled the actual realization and interpretation of such experiments. Physical possibility is thus to be distinguished from technological or epistemic possibility - from what happens to be technically achievable or cognitively imaginable by some particular historically situated community. Nevertheless, our judgments about which SPs are physically (im)possible are to be guided both by our theorizing and by our past experience, and as such are as fallible as any other empirical-scientific claim: superluminal signalling might be physically possible after all. Note that physical possibility is here not articulated in terms of ``laws of nature'', as in `phenomenon P is possible if it complies with laws of nature L'. That a SP is physically possible is taken as a primitive fact, as primitive as the fact that I ate breakfast today - a fact that may, but need not, be epistemically-pragmatically graspable in terms of what we nowadays call `laws of nature'.\footnote{The above properties that partly delimit the concept of physical possibility by no means constitute a complete philosophical reflection thereon, but only provide the preliminary material for one. That is, the commitment to the objectivity of physical possibility and to the ahistoricity of the truths that concern it, ought not to be taken dogmatically as the end of the story, but are to be elucidated in light of a broader reflection on the problematic notions of \textit{objectivity} and \textit{possibility}. Indeed, the very distinction between possible and impossible SPs is drawn within a larger space of possibilities, that we may impulsively denote as ``logical''. And this logical space is not in any way simply encountered in nature or imposed on us by a benevolent God, but involves some essential creative work and sweat on behalf of us, human beings. A proper understanding of physical possibility thus requires an adequate understanding of the origin and constitution of this logical space and of the distinctions drawn therein.
}

Getting back to Principle 2, that a PE-model should not posit absolutely indeterministic events has above been motivated by appealing to verifiability: if there is no experimental procedure that could possibly move us to accept that an event takes place with probability $p$ rather than $p'$ in phenomenon Z, then the very positing of the occurrence of this event is in some sense pathological - the logical empiricist would more radically say: meaningless. However, the connection between what there (possibly) is and what we can (possibly) know - or more generally, between truth and practice - is obviously too deep of an issue to be dogmatically pre-decided here. I would thus like to gesture towards another motivation for Principle 2, one that complements the above one, and that may possibly move even someone who is unmoved by verificationist-like considerations, in that it does not speak of the ideal possibility of verification but of the ideal \textit{possibility of reference}. That is, suppose that, according to a PE-model, a SP features an absolutely indeterministic event, meaning that there is a random event $\Omega$ succeeded by a random event $\Omega'$ in its causal future, but such that, for each $\omega^* \in \Omega$, there is no physically possible phenomenon wherein the former event takes deterministically value $\omega^*$ and wherein the latter events are distributed according to the corresponding conditional distribution $\mu(\omega'|\omega^*)$. The phenomenon in question, according to the model, thus features a statistical mixture of various mathematically representable but physically impossible phenomena - the impossible phenomena here each corresponding to the deterministic occurrence of some $\omega^* \in \Omega$. Even though each instance of the described phenomenon is assumed to also be an instance of phenomenon A or phenomenon B, neither A nor B are physically possible in the sense that there are no non-statistical conditions (referable to by some identifier) that would imply the deterministic taking place of either of them. Whereas I can point at a sequence of events and assert that either A or B occurred here, there is no possible sequence of events for which anyone could ever even ideally be licensed to assert that, say, A occurred there. The contact with the posited phenomena A and B is thus fundamentally limited  to the disjunction `A or B': nobody has nor ever will knowingly encounter phenomenon A. These ``semantic'' gestures are to the very least to mark an important difference between such speculatively posited phenomena and those phenomena that are only contingently indeterministic, and that are thus at least ideally referable to. But whether this difference - the ideal (im)possibility of reference - is sufficient as to remove once and for all absolute indeterminism from every healthy physical ontology is admittedly not entirely settled here, but would again require a more thorough reflection on the connections between truth and practice, or here, between the physical existence/occurrence of X and the ideal possibility of referring to an instance of X. We will however stop at the hereby induced note of uneasiness and henceforth tentatively assume that adequate PE-models do satisfy Principle 2.

Let us recapitulate. A PE-model is an adequate description of a SP $Z$ if its posited distribution over events in the spatiotemporal regions of relevance agrees with the statistics of these events as they actually unravel in instances of $Z$. Furthermore, the ideal possibilities of verification or reference seem to impose the validity of Principle 2 - that is, that no absolutely indeterministic event may be posited by a PE-model. Now we will list a few more observations on the properties of PE-models and of their relation to SPs. First, the spatiotemporally localized events thematized by PE-models are not restricted to be neither ``observable'' nor describable in terms of ``observational vocabulary'', and may thus range from the likes of audible signals produced by particle detectors to the presence of localized electrons with different spin orientations. One restriction that they however do need to abide by - and that they share with identifiers - is that the posited events need to be describable in non-statistical terms. That a drop of water evaporates with chance $p$ at location $x$ is thus not a valid event; similarly, albeit less explicitly, the presence of an electron in a quantum-mechanical state $\rho$ at location $x$ is also not an event, due to the quantum state having only statistical empirical significance.\footnote{These events have `only statistical empirical significance' in the sense that the adequacy of their positing is not ideally ascertainable for any particular event qua particular, but only qua an element of an ensemble. But why should we prohibit PE-models from positing such events? Are we sure we are not imposing an arbitrary and unreasoned prohibition here? Recall that every ensemble requires a non-statistical identifier that determines each element thereof as an element of that ensemble. The occurrence of an event with only statistical empirical significance - e.g. the presence of an electron in some quantum state at location $x$ - presupposes the taking place of another event that can be described in non-statistical terms - e.g. that an atom of a particular type emitted an electron in the past of $x$. What is distinct about such events is thus that their positing a priori presupposes the taking place of other events. The very idea of a priori independent spatiotemporally localized events seems to force us to consider only events describable in non-statistical terms. This gesturing however obviously requires further thought.} 
Second, note that the relationship between SPs and their adequate PE-models is many-to-many. Namely, any SP has many adequate PE-models already in virtue of the latter's redundancy: as previously mentioned, any two models that agree in their distribution over events located in the SP's region of interest are equally adequate. There is however a further factor that contributes to the plurality of adequate PE-models, which derives from the underdetermination of the size of the ``infinitesimal spatiotemporal regions'' that are coordinated to the model's elements of $\mathbb{R}^4$. One and the same SP may be described at levels corresponding to different spatiotemporal scales: the toss of a coin may be described in terms of hands, coins and tables, or in terms of molecules that these are constituted of. Moreover, any PE-model may adequately describe many different SPs - again, in virtue of its redundancy. This can be observed on the trivial case of a PE-model whose marginal distributions over events in two separated spatiotemporal regions can be used to describe two different unfair coin tosses and that thus furnishes an adequate model of two different SPs. Finally, note that in the same fashion in which many scientific models may generally share the same structure - think of the same mathematical forms appearing in physics and in economics - different PE-models may also share the same PE-structure, thereby making the connection between SPs and PE-structures even looser than the corresponding relation between SPs and PE-models.

\subsection*{III.I. Example} A plethora of concepts have so far been introduced, including SPs, their identifiers and regions of relevance, PE-structures, PE-models and conditions of adequacy. It is time to illustrate all of these on a simple exemplary SP. The SP will figure two machines, a machine of `type A' and a machine of `type B'. A machine of type A, when turned on, tosses a fair coin that lands a few seconds after, and depending on the outcome, emits either a ball of mass $m>1$kg or a ball of mass $m<1$kg. A machine of type B, when turned on, indicates with $90$\% efficiency whether the mass of any ball that hits it exceeds $1$kg or not, once it hits it. The SP's identifier reads as follows: \textit{A machine of type A and a machine of type B are placed at some distance facing each other, in such a way as for machine B to be hit by any ball that is emitted by machine A. The two machines are simultaneously turned on and are left to freely operate for a minute.} The region of relevance $X$, albeit vaguely defined due to the vagueness of the identifier, can at least be said to comprise a) the landing of the coin, that occurs, say, at $x$, b) the measurement of the flying ball's mass, that occurs at some future location, say, $y$, and c) the region connecting these two, wherein the ball travels between the machines. Note that the identifier accordingly refers only to events that occur in the past of $X$, specifying the initial conditions that determine the subsequent statistical unravelling of the thematized events.

The basic conditions for a PE-model to adequately describe our SP are for it not to posit the taking place of absolutely indeterministic events and for its posited distribution to agree with the one that is so-to-say given by the phenomenon. Events that any such adequate model can to the very least posit, and that are implicitly suggested by the SP's identifier, are the landing of the coin at location $x$ and the reading indicated by the second machine at $y$. Both of these random events are dichotomic: the coin can either land heads or tails and machine B indicates that the ball's mass is either heavier or lighter than $1$kg. The two events' associated sample spaces can thus be chosen as $\Omega_x \cong \Omega_y \cong\left\{0,1 \right\}$ and their statistical distribution can be represented by marginal distribution $\mu_{\left\{x,y \right\}}$ induced by PE-model $\mu$. A PE-model will thus adequately describe our SP only if the said marginal distribution corresponds to what actually unravels in instances of our SP. It is simple to inspect that the fairness of the coin and the efficiency of machine B imply that the distribution needs to satisfy 
\begin{equation*}
    \mu_{\left\{x,y \right\}}(c,m)=\frac{1}{2}\left(0.9\delta_{c,m}+0.1\delta_{c,m \oplus 1} \right),    
\end{equation*}
for all $c \in \Omega_x$ and $m \in \Omega_y$.

Note that our PE-model contains severe redundancy, and is accordingly equally severely underdetermined, in that the distribution of its posited random events outside of the region of relevance do not matter for the adequacy of the model. Every PE-model however also asserts the taking place of certain events within the region of relevance besides the two aforementioned events, which accordingly do matter for the model's adequacy. For example, an adequate PE-model may track the ball along its trajectory by positing that at each spatiotemporal location $z$ - positioned somewhere along the ball's trajectory between $x$ and $y$ - a ball is present whose mass either exceeds or does not exceed $1$kg. Representing these two possible events at some location $z$ by $\Omega_z \cong \left\{0,1 \right\}$, any adequate PE-model should trivially satisfy 
\begin{equation*}
  \mu_{\left\{x,y,z \right\}}(c,m,m_z)=\frac{1}{2}\left(0.9\delta_{c,m}+0.1\delta_{c,m \oplus 1} \right)\delta_{c,m_z}  
\end{equation*}
for all $m_z \in \Omega_z$, meaning essentially that the ball does not change its mass during its flight. Note that this positing also abides by Principle 2, in that there obviously is a physically possible SP in which a ball of mass lighter than $1$kg deterministically travels from $z$ to $y$ and elicits distribution 
\begin{equation*}
    \mu_{\left\{y,z \right\}}(m,m_z)=\left(0.9\delta_{0,m}+0.1\delta_{1,m} \right)\delta_{0,m_z}
\end{equation*}
and analogously so for the case of the ball being heavier than $1$kg. The indeterminism of the random event at $z$ is thus contingent on events in its causal past. In fact, this indeterminism is not only contingent but also reducible, in that the conditions that determine, say, $m_z=0$ are in fact instantiated in the very phenomenon that we are describing, so that the indeterminism can be interpreted as deriving from a contingent lack of control over an otherwise deterministic unravelling - here, a lack of control over the outcome of the coin toss.

\section*{IV. PE-models of Quantum Statistical Phenomena}
The variety of SPs found studied across the natural and social sciences includes also \textit{quantum} SPs - that is, those that are nowadays canonically described using quantum-mechanical models (QM-models). As mentioned at the outset, the problematics of quantum SPs, which makes them deserve a treatment on their own, derives from a lack of consensus about what can plausibly be said to go on, let alone on what ``really goes on'', in any of their instances. Take a simple Stern-Gerlach experiment, whereby an electron is sent through an inhomogeneous magnetic field before impinging on a detector. Even though there is practically absolute agreement on which QM-model is to be used to account for this phenomenon, there is no consensus neither on which objects figure in the phenomenon nor on which events take place therein, besides the pre-theoretically agreed upon presence of macroscopic instruments and changes that they undergo. Relatively popular proposals range from the quasi-classical-mechanical conviction that there is a localized particle that travels continuously between the instruments, all the way to the Bohrian conviction that there is no healthy sense in speaking of what happens in between the emission and the detection of the electron, at least not without a dose of relativization to some sort of context. There is consequently a wide and non-easily-traversable gap between a QM-model of a particular SP and any candidate PE-model that purports to describe that same SP. Our goal here is to inspect some properties of the translation manual that transposes QM-models of quantum SPs into PE-models that are adequate descriptions thereof according to the minimal criteria of adequacy sketched beforehand. 

Take some quantum SP - e.g. the aforementioned Stern-Gerlach experiment, the EPR experiment, or some experiment on high energy protonic beams. Which PE-models are adequate descriptions of these phenomena? The answer is to be attuned to the best theoretical accounts that we already possess of these SPs, and that are provided by QM-models. The problem however, as mentioned above, is that these say close to nothing about which events take place in the experiments. In fact, the use of a typical QM-model presumes a prior ``naive'' positing of events assumed to occur in the experiment, and whose statistical distribution is to be correctly embedded within an abstract mathematical model - these events usually taken to correspond to the likes of the setting-up of experimental equipment and outcomes of measurements, the stuff that information-theorists denote by ``inputs'' and ``outputs''. The distribution of the remaining events however remains unthematized, making it unclear whether any other events can legitimately be said to take place at all. More precisely, quantum SPs, as all other SPs, have identifiers and more or less vaguely defined regions of relevance $X \in \mathcal{X}$ - the identifiers usually referring to regions in the causal past of $X$, e.g. specifying the quantum state that a quantum system is initially in. The use of a QM-model in the describing of a quantum SP, as already stated, presumes a division among the latter's regions of relevance $X$ into those in which certain measurements occur - call them $X_m$ - and those in which no measurement occur - $\overline{X}_m$. For example, the Stern-Gerlach experiment is analyzed into the regions in which the electron is travelling freely, and the region in which the electron's spin is being measured; or take a high-energy experiment at an accelerator, whose regions of relevance are subdivided into those in which the particle beams are travelling through the accelerator, and those in which certain particles are recorded by detectors. Now, it is clear that any PE-model, if it is to adequately describe a quantum SP, ought to the very least reproduce the correct distribution $P_m$ of the naively posited events in $X_m$, by embedding it within a distribution $\mu$, whose corresponding marginal $\mu_{X_m}$ is to coincide with $P_m$. It remains however unclear what is the relationship that obtains between the QM-model and the distribution of events in $\overline{X}_m$, in that the model seemingly does not presume the taking place of any such events at all.

Nevertheless, there is a reason why a non-empty collection of regions $\overline{X}_m$ is included among the regions of relevance $X$, in addition to the seemingly unproblematic regions $X_m$. One way of seeing this is by noting that many different QM-models may be used to reproduce one and the same distribution $P_m$ of the measurement outcomes in $X_m$ in any particular quantum SP. Nevertheless, it would obviously be wrong to say that all of these models are for that reason equally adequate: even though the statistical distribution of measurement outcomes in a Stern-Gerlach experiment may be reproduced by a model that posits a traveling atom instead of an electron, even the most radical empiricist would agree that the atomic model would be at fault here - indeed, if we were to inspect for which particle is present in the experiment, we would find an electron, and not an atom (or at least it would appear \textit{as if} it were so). The adequacy of QM-models therefore does suggest more than the mere occurrence of measurements in $X_m$. The problem, however, is to state which events exactly can be said to occur in regions $\overline{X}_m$, such for it to be in accord with our basic desiderata of adequacy of PE-models. If we are to make progress here, we are first to recognize - as the previously listed examples already indicate - that the class of quantum SPs is widely heterogeneous, and that this heterogeneity is consequential for any assessment of PE-models' adequacy as descriptions thereof. Let us thus start by roughly classifying quantum SPs into the following possibly non-exhaustive categories: phenomena involving (a)  single localized quantum systems, (b) multiple localized quantum systems, (c) multiple non-localized quantum systems, (d) quantum fields. In what follows we will only survey some properties of phenomena and their models that fall under categories (a) and (b), leaving the other ones for a future treatment. 

\subsection*{IV.I. Single Localized Quantum Systems}
By `phenomena involving single localized quantum systems' I intend those phenomena whose canonical QM-models posit a single non-relativistic ``system'' that can be associated relatively accurately to a spatiotemporal trajectory $u: \left[0,1\right] \rightarrow \mathbb{R}^4$. An example thereof is a quantum-optical experiment in which a photon is constrained to travel through a waveguide along which various devices are placed that are intended to measure some internal degree of freedom pertaining to the photon (e.g. its angular momentum). For each spatiotemporal location $u(t)$ along the system's trajectory, the QM-model - here conventionally presented in the Schrödinger picture - posits a quantum state $\rho(t)$, whereby the relation between the states posited at different locations is determined either by a unitary transformation or by some more general completely positive map. The model also posits a list of positive operator-valued measures (POVMs) $\left(\left\{\Pi^{(1)}_{a_1} \right\}_{a_1},...,\left\{\Pi^{(n)}_{a_n} \right\}_{a_n}\right)$ that, together with the quantum states, determine the distribution of outcomes $P(a_1,...,a_n)$ of the measurements that are performed in the experiment at locations $u(t_1),...,u(t_n)$. In previously introduced terms, the experiment's regions of relevance are partitioned into those regions that host measurement outcomes - here $X_m=\left\{u(t_1),...,u(t_n) \right\}$ - and the remaining regions - here $\overline{X}_m=\text{Im}(u) \setminus X_m$, where Im($u$) is the image of function $u$. The latter regions $\overline{X}_m$, albeit canonically not associated to measurement outcomes, are nevertheless assumed to host a photon to which quantum-mechanical states are accordingly associated.  

Let us now inspect some properties of PE-models that appear to adequately describe experiments of the above introduced type. Here we will focus only on a subtype of such experiments, which will nevertheless enable us to draw some interesting lessons. The said subtype contains experiments in which (a) the presumed measurements that take place at $T\equiv \left\{u(t_1),...,u(t_n) \right\}$ can be modelled by \textit{projection}-valued measures (PVMs) $\mathbf{\Pi}\equiv \left\{\left\{\Pi^{(1)}_{a_1} \right\}_{a_1},...,\left\{\Pi^{(n)}_{a_n} \right\}_{a_n}\right\}$, and in which (b) the quantum system is presumed to be \textit{closed} in between consecutive measurements, its dynamics thus being canonically describable by a family of unitary operators $U(t)$. A QM-model $Q_E$ of an experiment $\mathcal{E}$ of this subtype is accordingly given by the list $Q_E=\left(\rho,U(t),T,\mathbf{\Pi} \right)$, where 
\begin{itemize}
    \item $\rho$ denotes the quantum state that the system is in at location $u(0)$,
    \item $U(t)$ are the unitary operators that specify the dynamics in between the measurements, and that satisfy $U(t)U(t')=U(t+t')$,
    \item $T$ are the spatiotemporal locations at which the measurements occur, 
    \item $\mathbf{\Pi}$ are the PVMs that model the latter measurements.
\end{itemize}

Quantum theory implies that the probability that an execution of $\mathcal{E}$ results in outcomes $(a_1,...,a_n)$ is given by

\begin{equation}
    P(a_1,...a_n)=\Tr\left[\Pi^{(n)}_{a_n}(t_n)...\Pi^{(1)}_{a_1}(t_1)\rho \Pi^{(1)}_{a_1}(t_1)...\Pi^{(n)}_{a_n}(t_n) \right],
\end{equation}
where $\Pi^{(i)}_{a_i}(t)\equiv U^{\dagger}(t)\Pi^{(i)}_{a_i}U(t)$. Now, one of the minimal conditions for a PE-model $\mu$ to be an adequate description of $\mathcal{E}$ is for it not to contradict the above quantum-mechanically informed distribution, i.e. $\mu_{X_m}(a_1,...,a_n)=P(a_1,...,a_n)$. PE-models however also posit random events that take place at locations in between the events that correspond to measurements - i.e at each location in $\overline{X}_m$ - and this positing may or may not jeopardize their adequacy. Take for instance a simple PE-model that asserts that at each location in $\overline{X}_m$ an event occurs deterministically that corresponds to ``the presence of the photon at that location'': the adequacy of this model follows immediately in that, were one to check whether there is a photon at any location in $\overline{X}_m$, one ideally would indeed find a photon present there deterministically.\footnote{Let there be no misunderstanding: the `checking' referred to here is much more convoluted than checking, say, whether there is some milk left in the fridge, in that it requires surgically designed instruments and a surgically designed account of how we are to interpret the behavior of these instruments. This point of course requires a more detailed discussion, that might as well turn out quite consequential for some of the details in this section.} 

Quantum theory however appears to also allude to more substantial PE-models. In order to see this, consider \textit{another} experiment $\mathcal{E}^*$ whose corresponding QM-model is given by $Q_{E^*}=\left(\rho,U(t),T^*,\mathbf{\Pi}^* \right)$, where 

\begin{align*}
    &T^*=\left\{u(t^*_1),...,u(t^*_m)\right\}\supset T\\
    &\mathbf{\Pi}^*\equiv \left\{\left\{\Pi^{(1)*}_{a_1} \right\}_{a_1},...,\left\{\Pi^{(m)^*}_{a_m} \right\}_{a_m}\right\} \supset \mathbf{\Pi}
\end{align*}
and such that the marginal distribution on the events in $X_m$ coincides with the one given in experiment $\mathcal{E}$:

\begin{align}\label{compl cond}
    \sum_{\substack{a_{j_1},...,a_{j_{m-n}} \\ t^*_{j_k} \notin \left\{t_1,...,t_n \right\}}} P^*(a_1,...,a_m)&=\sum_{\substack{a_{j_1},...,a_{j_{m-n}} \\ t^*_{j_k} \notin \left\{t_1,...,t_n \right\}}} \Tr\left[\Pi^{(m)*}_{a_m}(t^*_m)...\Pi^{(1)*}_{a_1}(t^*_1)\rho \Pi^{(1)*}_{a_1}(t^*_1)...\Pi^{(m)}_{a_m}(t^*_m) \right]\\
    &=P(a_1,...a_n).
\end{align}

$\mathcal{E}^*$ differs from $\mathcal{E}$ only in featuring measurements that are not actually performed in $\mathcal{E}$. These measurements however could be performed, and if they were performed, their outcomes would be distributed in accord with what QM-model $Q_{E^*}$ implies. Recall that PE-models may freely posit ``theoretical'' events that do not correspond neither to measurement outcomes nor to ``observable'' events, as long as this positing has the correct empirical consequences and as long as it complies with Principle 2 in not speculating on the taking place of absolutely indeterministic events. Having this in mind, consider a PE-model $\mu^*$ that asserts that at locations $T^*=(u(t_1^*),...,u(t_m^*))$ random events occur that take values in $\Omega_{u(t_1^*)},...,\Omega_{u(t_m^*)}$, where the elements of $\Omega_{u(t_i^*)}$ stand for events that correspond to the presence at $u(t_i^*)$ of the quantum system in question ``possessing'' a definite value $a_i$ of the magnitude (that can be said to be) measured by the corresponding PVM $\left\{ \Pi^{(i)*}_{a_i}\right\}_{a_i}$. Suppose furthermore that the marginal distribution $\mu^*_{T^*}$ coincides with distribution $P^*(a_1,...,a_m)$ that is implied by the quantum-mechanical model $Q_{E^*}$. It follows that $\mu^*$ is an adequate PE-model not only of experiment $\mathcal{E}^*$, but also of experiment $\mathcal{E}$, since (a) the empirical consequences of its theoretical posits comply with quantum theory, which in turn complies with empirical evidence, and (b) none of the posited random events is absolutely indeterministic, in that, for each $i$ and $a_i$, there is a physically possible experiment $\mathcal{E}^{(i)}_{a_i}$, whose random events in the future of $u(t_i)^*$ are distributed according to the conditional distribution $P^*(a_m,...,a_{i+1}|a_i)$. 

Therefore, for any pair of experiments $\mathcal{E}$ and $\mathcal{E^*}$ that are of the above subtype and that satisfy condition \eqref{compl cond}, any PE-model that is adequate of $\mathcal{E^*}$ is also adequate of $\mathcal{E}$. Let us now illustrate this on a simple experiment involving a single spin-measurement of an electron. Take $Q_E=\left(\rho,U(t),T,\mathbf{\Pi} \right)$, with 

\begin{itemize}
    \item $\rho=\ket{0}_z\bra{0}$,
    \item $U(t)=e^{-i \sigma_z t}$
    \item $T=\left\{ u(1) \right\}$
    \item $\mathbf{\Pi}= \left\{\ket{a}_x\bra{a} \right\}_{a=0,1}$,
\end{itemize}

where $\ket{0}_z$ is the state in which the electron's spin is oriented along a conventionally chosen $z$-axis, $\sigma_z$ is the standard Pauli operator, and $\ket{a}_x=\frac{1}{\sqrt{2}}(\ket{0}_z+(-)^a\ket{1}_z)$. The experiment features only one ``$x$-measurement'', whose outcomes are, according to quantum theory, distributed as $P(a)=\frac{1}{2}$. Now consider another experiment $\mathcal{E^*}$ that differs from the former one only in including a further ``$z$-measurement'' at some time $0<t<1$, such that its QM-model is given by $Q_{E^*}=\left(\rho,U(t),T^*,\mathbf{\Pi}^* \right)$, with $T^*=(u(t),u(1))$, and $\mathbf{\Pi}^*= \left\{ \left\{\ket{a'}_z\bra{a'} \right\}_{a'}, \left\{\ket{a}_x\bra{a} \right\}_{a} \right\}$. Quantum theory implies that the two measurement outcomes in $\mathcal{E}^*$ are distributed according to $P^*(a',a)=\frac{1}{2}\delta_{a',0}$, whose marginal distribution $P^*(a)$ thus agrees with the former distribution $P(a)$. Now consider a PE-model $\mu^*$ that posits that at locations $u(t)$ and $u(1)$ an electron is present whose spin is oriented respectively parallelly or anti-parallelly to the $z$-axis and parallelly or anti-parallelly to the $x$-axis, such that the marginal distribution $\mu^*_{\left\{u(t),u(1)\right\}}(a',a)$ agrees with $P^*(a',a)$. It follows from the above discussion that $\mu^*$ is an adequate PE-model of $\mathcal{E}$. 

Another PE-model of $\mathcal{E}$ can be analogously constructed by considering a third experiment $\mathcal{E}^{**}$ that differs from $\mathcal{E}^*$ only in featuring an intermediate $x$-measurement, instead of a $z$-measurement. That is, the QM-model of $\mathcal{E}^{**}$ is given by $Q_{E^{**}}=\left(\rho,U(t),T^*,\mathbf{\Pi}^{**} \right)$, where $\mathbf{\Pi}^{**}= \left\{ \left\{\ket{a'}_x\bra{a'} \right\}_{a'}, \left\{\ket{a}_x\bra{a} \right\}_{a} \right\}$. Quantum theory now implies that the two outcomes are distributed as 
\begin{equation*}
    P^{**}(a',a)=\frac{1}{8}|e^{-i(1-t)}+(-1)^{a+a'}e^{i(1-t)} |^2, 
\end{equation*}
which again induces the same marginal distribution $P^{**}(a)=\frac{1}{2}$. A PE-model $\mu^{**}$ that posits that at locations $u(t)$ and $u(1)$ an electron is present whose spin is at both locations oriented parallelly or anti-parallelly to the $x$-axis, and whose distribution $\mu^{**}_{\left\{u(t),u(1)\right\}}(a',a)$ agrees with $P^{**}(a',a)$, is again an adequate model of our original experiment $\mathcal{E}$. \textit{Two different PE-models}, $\mu^*$ and $\mu^{**}$, thus \textit{turn out to be adequate of one and the same phenomenon}. 

As stated beforehand, that a SP admits of more PE-models is not surprising per se, as long as the models' assertions do not contradict each other. After all, stated somewhat vaguely, different models may be thought of as (i) capturing different ``aspects'', or (ii) describing different ``levels'' of one and the same phenomenon. Even models of a simple coin toss exhibit both of these sorts of plurality: (i) one PE-model may describe the linear momentum of the coin along its trajectory, whereas another may refer only to its angular momentum, or (ii) one PE-model may refer to hands and coins, whereas another to microscopic molecules that the hands and coins are purportedly constituted of. Nevertheless, in both of these cases, the different models can be conjoined into a single model that e.g. speaks both of the molecules' linear and angular momenta \textit{and} of coins and hands. The plurality of adequate PE-models of a single phenomenon may thereby be interpreted as sourced from the plurality of our contingent interests in this-or-that aspect or this-or-that level of one and the same unitary phenomenon. The plurality that engenders our quantum-mechanical example is however significantly different, in that the different models can notably \textit{not} be conjoined into a single model - one that would, in our particular case, assert that at location $u(t)$ an electron is present whose spin is aligned both parallelly or antiparallelly to the $z$-axis \textit{and} parallelly or antiparallelly to the $x$-axis. The particular impossibility stated here is certainly not logico-mathematical, in that one may freely construct a consistent PE-model $\Tilde{\mu}$ that asserts the taking place of such joint events and whose marginal distributions agree with what $\mu^*$ and $\mu^{**}$ assert.\footnote{An example of such a joint model is given by a distribution $\Tilde{\mu}$ that satisfies $\Tilde{\mu}\left[\left(a_z,a_x \right),a\right]=\frac{1}{\mu^*\left(
a\right)}\mu^*\left(
a_z\right)\mu^{**}\left(
a_x\right)\mu^*\left(
a|a_z\right)\mu^{**}\left(
a|a_x\right)$, where the subscripts are left implicit for the sake of legibility.} It is instead a physical impossibility, our epistemological contact with which is mediated by theoretical and experimental physics in their current state. Indeed, according to contemporary physics, the hypothetically proposed model $\Tilde{\mu}$ would \textit{not} be an adequate description of our experiment, since it would violate Principle 2 by positing the taking place of necessarily indeterministic events - as there is no evidence that an experiment is possible in which a single electron is prepared in a state that can plausibly be said to correspond to its spin having definite alignments relative to different spatial axis.

The above lesson concerning the irreducibility of the plurality of adequate PE-models of our experiment $\mathcal{E}$ straightforwardly generalizes to all experiments on single localized quantum systems, in so far as they all, according to their canonical QM-models, feature quantum systems to which a plurality of different magnitudes can all synchronically be associated, but not all of these magnitudes can also synchronically be specified to take definite values - which implies that most, though not all, such quantum SPs admit an irreducible plurality of adequate PE-models.\footnote{An example of a quantum SP that does \textit{not} admit of complementary models is one whose associated QM-model is given by $Q=\left(\rho,U(t),T,\mathbf{\Pi} \right)$, with $\rho=\ket{0}_z\bra{0}$, $U(t)=e^{-i \sigma_z t}$, $T=\left\{ u(1) \right\}$, and $\mathbf{\Pi}= \left\{\ket{a}_z\bra{a} \right\}$, whereby the measurement outcome is distributed deterministically as $P(a)=\delta_{a,0}$. It is simple to notice that the positing of any event that corresponds to a magnitude that is incompatible with $z$-spin - e.g. the positing of events corresponding to $x$-spin at some location $u(t)$ - would not induce the same marginal distribution $P(a)$.} This lesson can be interpreted as a manifestation, at the level of PE-models, of something akin to the \textit{complementarity} that Niels Bohr regimented as one of the basic marks of quantum-mechanical phenomena: different PE-models offer seemingly irreducibly complementary descriptions of one and the same quantum SP. Upon asking ``Which events occur in a quantum SP?'' the answer we got was non-unique, and this non-uniqueness appears not to be explainable away by the contingencies of our all-too-human capabilities and interests. What should we make of this? Is the situation we ended up in somehow problematic, and if so, how are we to resolve it? 

\subsection*{IV.II. Coping with Complementarity}

In order to better understand our own situation, let me trace the thread that has brought us here, the micro-history of our micro-investigation, so to say. At the beginning we faced the broad category of SPs that pop up all across our scientific and non-scientific lives. Among these we tentatively included quantum phenomena, phenomena that happen to currently be accounted for with the use of hermeneutically opaque QM-models, but that do seem to exhibit regularities similar to those that are exhibited by roulette spins, coin tosses and drug-induced patients' recoveries. Then, guided by the general conviction that SPs can be dissected into spatiotemporally localized random events, we mathematically rectified our ontological category - we prepared it for a more rigorous scientific treatment - by turning its members into correlates of a certain class of mathematical models, our PE-models. We thereby introduced some basic conditions of adequacy that are to underlie our evaluation of a PE-model as a description-representation of a given SP, providing us with criteria that are to guide our trust into the answers that a PE-model suggests to our elementary question: ``Which events occur in this SP?'' Finally, we forcefully transported the same conceptual-mathematical machinery into the quantum realm in order to find an event-ontology of quantum phenomena, which has however pushed us into a curious situation: our algorithm tells us that a plurality of non-logically-contradictory but non-conjoinable PE-models are all adequate descriptions of the same quantum SP. And it is at this juncture that we are forced to re-think our own situation.

It might be of some use to compare the above with the following fictional analogy. A storm is occurring on a distant planet and a group of scientists wants to learn something about its audio-visual outlook. The scientists design various cameras that appear to work reliably here on Earth - they work well when confronted with our familiar storms - and they send a collection of them to the distant planet. The collection is so chosen as to consist of (i) some cameras that record both color and sound, (ii) some that record sound but no color, and (iii) some that conversely record color but no sound. Upon receiving their desired videos and thoroughly analyzing them, the scientists spot that the recordings within category (i) contradict each other, by e.g. depicting differently the color of an avalanche of sand scattered by the wind or disagreeing on the sound produced by this scattering. On the other hand, the recordings within category (ii) all agree on the sounds that they registered, and same for the videos in (iii), which all agree on the recorded colors. After generations of trials, methods, groups and efforts, a broad scientific consensus is reached that there is no physically possible method that could produce a reliable representation that integrates both the audial and the visual aspects of the storm - despite the fact that such an integration can be synthesized on a computer. What are we to make of this? Are the peculiarities to be traced to unknown mulfunctionings of our scientists' cameras, to the contingencies of their technological limitations? Or is the storm a quasi-Lovecraftian phenomenon whose translation into ordinary audio-visual representations differs radically from such translations of our ordinary storms here on Earth? 

Getting back to our own situation: PE-models were originally designed to provide event-ontological descriptions of SPs, but were so designed on the basis of our familiar contact with ordinary SPs. We tentatively tried to employ their use in the non-ordinary, in the quantum, and we found out that this stretching of our descriptive instrument resulted in a peculiarity. An instrument was born in the familiar, it was pushed into the non-familiar, and it seems to have displayed radically non-familiar behavior. We ought to re-evaluate our relation to our instrument, our trust in what the instrument appears to be telling us. In fact, the irreducible plurality-complementarity that we found in our confrontation with the quantum does not in the first instance concern the SPs per se, but only their adequate PE-models: the videos-instruments are in the first instance plural, not the storm itself. That is, the validity-adequacy-appropriateness of a description - with the criteria of adequacy springing from the ordinary - does not imply the truth of that description, not without side commitments concerning the relation between the description and the described or side modifications-mutilations of our conception of truth. In what follows we will thus critically reflect on the bearing that the complementarity of PE-models is to have on a possible event-ontology of quantum phenomena. The reflection will be only preliminary, and mostly negative, in that it will point out difficulties that arise with some relatively straightforward attempts at digesting the latter bearing, leaving us at the kind of impasse that is by now usual in encounters between ontology and quantum theory.

Schematically, there are three different ways of ontologically digesting the irreducible plurality of mutually complementary adequate PE-models of quantum SPs. One can take ontologically seriously \textit{all}, \textit{some} or \textit{none} of the models that partake in the plurality - whereby to take ontologically seriously a PE-model of a SP is to contend that the events referred to in the model really take place in that SP, in the very same sense in which a coin is said to really land heads or tails in any of its tosses. If one brutely takes \textit{all} adequate PE-models ontologically seriously, then, due to there being irreducibly complementary adequate PE-models, one is forced to admit the occurrence of events that are not referred to in any adequate PE-model (e.g. that at location $u(t)$ an electron is present whose spin is aligned parallelly/antiparallelly to the $x$-axis \textit{and also} to the $z$-axis). This amounts to committing oneself to a ``hidden-variable (HV) model'' that integrates the plurality of complementary PE-models only by positing events whose empirical and ontological significance is unclear at best, in that they do not comply with Principle 2. 

Note that this criticism of attempts at resolving the quantum problematic by appealing to HV models is not based on the theorems proven by Bell, Kochen, Specker and others, which prove logico-mathematically that no HV theory that satisfies a list of theoretical desiderata can reproduce the empirical predictions of quantum theory. The point advanced here is rather more primitive and thus applies more extensively: even those HV models that do satisfy some internal theoretical desiderata and that reproduce the statistical distribution of ``inputs'' and ``outputs'' in some particular quantum SPs are nevertheless pathological descriptions of those SPs, due to their speculation on occurrences of absolutely indeterministic events. This is of course so for those models that posit faceless events, that is, mere mathematically represented \textit{placeholders} of possible events, such as it would be the case for an application of the Spekkens toy model of a qubit in the modelling of an experiment on an electron's spin (Spekkens, 2007), or an application of the Werner-model in the account of an EPR-like experiment that does not violate Bell's inequalities (Werner, 1989): the posited hidden events are left completely unspecified-faceless, and unsurprisingly so, as there is currently no empirical evidence that could guide their specification.\footnote{This is by no means a criticism of the great value these models have for our understanding of the structure of quantum (information) theory, but only of the hypothetical temptation to take them ontologically seriously.} But the pathology is by the way inherited also by those HV models that do not posit faceless events - whose events are so-to-say materially filled - such as by a De Broglie-Bohmian model of the double-slit experiment, whose posited events indicate the seemingly unproblematic presences and absences of material particles in infinitesimal spatiotemporal regions. The pathology of this positing however derives again from a violation of Principle 2, in that the particle trajectories posited by the model are asserted to be physically realizable only under the condition that they are statistically randomized. Stated somewhat provocatively, the double-slit experiment is thereby conceived as a statistical mixture of physically impossible phenomena featuring physically impossible particle trajectories.\footnote{Here is an articulation of the provocative insinuation above that echoes the discussion in Section III: the De Broglie-Bohmian trajectories are `physically impossible' in the sense that there is no collection of ideally verifiable non-statistical conditions - referable to by some identifier - that ideally imply the obtaining of said trajectories. Whereas a proponent of this theory may of course keep asserting that these trajectories are physically possible, and that exactly one out of a plurality $\left\{x_1(t),...,x_n(t)\right\}$ of them is instantiated \textit{here} in \textit{this} particular sequence of happenings, there is no physically possible sequence of happenings that she could ever possibly point to and assert that they instantiate exactly, say, trajectory $x_i(t)$. The posited trajectories have never been nor will ever be individually referable to.} 
This impossibility may perhaps be relaxed by e.g. taking the Born rule as physically contingent on the history of the universe, which would turn a pathological theory into a potentially non-pathological one only by transforming it into an empirical contender with the currently evidentially supported quantum theory.\footnote{This meritable suggestion is most prominently being developed by Anthony Valentini; see e.g. Valentini (2023).}

The second digestive strategy mentioned above severs the irreducible plurality of PE-models by imbuing only some of them with ontological significance. For instance, one might take ontologically seriously only those models that posit events that correspond to measurements whose PVM elements commute with the momentary quantum state, or with the momentary Hamiltonian operator that generates the free development of the thematized system.\footnote{These proposals would be akin to event-ontological renditions of some of the object-property-ontological proposals that fall under the broad category of ``modal interpretations'' (see Lombardi \& Dieks, 2021).} The problem with this class of proposals is however its seeming arbitrariness. Even though one can logically-consistently reduce the plurality of models by focusing only on a subclass thereof, this reduction ought to be argued for. There is no legitimate reason to take, say, the mute and colored recordings as more representative of the actual storm than the other recordings and to thereby conclude that there are in fact no noises produced by the storm. Any judgment that is to be made concerning the relevance of the videos - or in our case: the relevance of the PE-models - ought to be an egalitarian judgment. 

The third and final class of proposals rejects all of the mutually complementary PE-models as guides to a possible event-ontology of quantum SPs: none of these PE-models can be said to accurately state what really goes on in a quantum SP - be it due to (a) these models not being fit for the job, or (b) the job itself being somehow misguided. I take a prominent rendition of something close to view (a) to be extractable from Richard Healey (2017)'s pragmatist-inferentialist account of quantum theory, that appeals to \textit{decoherence} as the arbiter of the occurrence of events: decoherence selects a mutually compatible collection of magnitudes that can all synchronically take determinate values, thereby avoiding the irreducible plurality of the kind manifested in the positing of events to spatiotemporal regions that host negligible decoherence.\footnote{Let me emphasize that what is written here is what I take to be tentatively implied by Healey's account of quantum theory, and not what he explicitly endorses: in fact, his book is very clear on avoiding speculations on a possible ``quantum ontology''.} As I see it, the problem with this kind of proposal is once again its seeming arbitrariness: whereas decoherence certainly can do the job of severing the plurality of PE-models, why should we follow its advice? What makes it the adequate arbiter here? After all, positing non-trivial events to decoherence-free spatiotemporal regions does bring with it inferential commitments, and thus seems to be meaningful even according to Healey's inferentialist conception of the meaningfulness of empirical statements: by asserting that a particle with angular momentum $l$ is present in a decoherence-free region with probability $p$, we are, among other things, committing ourselves to what would occur if we intervened on the said region, by e.g. placing a measurement device there or turning on a magnetic field. Another potential problem with a decoherence-relative criterion for ontological significance, one that is correlated to the metasemantic commitment to inferentialism, is its vagueness - i.e. how much decoherence? - implying that some assertions concerning event-occurrences somehow have more content than others. This is to the very least to induce some discomfort for an ontologist who aims to understand these statements as reporting something about the physical world around her rather than only embodying a mechanism for a better-or-worse coping with this same world.\footnote{The relationship between semantics and ontology, between reporting and coping, between truth and practice, the very meaning of claims about what there is and what occurs - none of these can nor should be settled in a few lines here. The discomfort I am referring to above is for now nothing more than that, an itch that needs to be addressed, and that may as well allow for a satisfactory resolution.}

On the other hand, option (b), the one that criticizes the very attempt of providing an event-ontology to quantum SPs - at least one of the kind that may possibly be provided by PE-models - may tentatively be extracted from the ``QBist'' and the ``relational'' accounts of quantum theory.\footnote{A recent presentation of the main tenets momentarily advocated by the proponents of QBism can be found in Fuchs (2023). For the relational interpretation, see for instance Di Biagio \& Rovelli (2022) and the updated version thereof in Adlam \& Rovelli (2023). The recent perspectivalist suggestion due to Dennis Dieks (2022) might also fit well among these. For a more general appraisal of perspectivalist-like treatments of quantum theory, see Adlam (2024).} According to these views, occurrences of events are somehow to be relativized either to agents-observers-subjects, to physical objects-systems, or more generally to ``perspectives'' of some sort. So talking about event-occurrences \textit{simpliciter} is misguided, except for particular quasi-classical situations, likely again identifiable by reference to decoherence. Now, none of these proposals has yet been developed into a fully self-enclosed account, so it might be premature to criticize them. I only want to highlight one general but hermeneutically essential point: in aiming to relativize the occurrence of all events to a certain X, we need to do our best to adequately elucidate the relation of these relativized ``occurrences'' to what we ordinarily mean by event-occurrences in the rest of science and in our everydayness. Hermeneutic opaqueness is otherwise to remain looming over these suggestions until the relationship is clarified between the purportedly relativized occurrence of radioactive decays and particle detections one the one hand, and the ordinary seemingly non-relativized occurrence of coin tosses, birthdays and molecular reactions on the other. Again, this is by no means intended as a destructive critique of these proposals, but an emphasis on the necessity of a further philosophical articulation thereof. And again, by philosophical articulation I do not mean a ``mere'' mathematical formalization of these theories - whose technical precision might as well obfuscate an underlying lack of philosophical rigor - but a reflection on the very intelligibility of relativized event-occurrences, one along the lines pursued e.g. by contemporary developers of a ``perspectival realism'' (see e.g. Massimi, 2022).

This does it for our brief inspection of the three schematic ways of coping with complementarity. What we have is definitely not a rigorous trilemma that takes us along three disjoint and exhaustive roads, proving once and for all the presence of irresolvable problems encountered at each of them. It is rather a cursory overview of vaguely delineated ways, that points out some of their obstructions, without proclaiming a final judgment on their absolute (in)adequacy. Let me say something positive now, albeit minimal, on what I think is a preliminary conclusion of this brief reflection. For starts, as already stated, we should be egalitarian in our evaluations of the various complementary models-videos, unless we find good empirically motivated reasons that are to pick out an ontologically preferred subclass thereof. So it does seem just that all of the mutually complementary adequate PE-models of a quantum SP are either all to be taken as equally adequate or as equally inadequate accounts of which events take place in that phenomenon. And given that I take a non-relativized event-ontological investigation to be \textit{prima facie} intelligible and thus worthy of being pushed to its ideal limits, together with the doubts I have expressed concerning the ontological significance of decoherence, I am lead to tentatively reconsider the first option, the one that attempts to take all of the complementary PE-models equally seriously, albeit now via a rendition thereof that attempts to avoid positing absolutely indeterministic events.

Again, that descriptions are plural and that we are free to roam among them is a platitude of both our theoretical and our non-theoretical lives; the significance of the quantum is however the aforementioned \textit{incompatibility} among these descriptions. But what is the sense of incompatibility that is at play here? Well, as already mentioned, it is obviously not logico-mathematical incompatibility, but nor is it the usual kind of physical incompatibility in the sense in which me being here is physically incompatible with me simultaneously being there: this usual sense is in fact incorporated \textit{within} a PE-model, in that the events that are posited to possibly occur at some location are assumed not to possibly all co-occur at that same location (e.g. an electron's spin cannot simultaneously point at direction $z$ \textit{and} at direction $-z$). Rather, two different descriptions $D_1$ and $D_2$ that are based on two complementary PE-models are incompatible in the sense that, although there may exist a SP that makes adequate both of these, there is no phenomenon that could make adequate a unified description that implies both $D_1$ and $D_2$. But what is the precise difference between a SP making adequate two descriptions $D_1$ and $D_2$, on the one hand, and a SP making adequate a unified description that implies $D_1$ and $D_2$, on the other? Does the former to the very least not entail the latter?

Take that $D_1$ and $D_2$ agree that in some particular spatiotemporal region of a thematized SP an electron is present whose spin points at some determinate direction, but that they disagree on the range of possible such directions: $D_1$ posits that the spin is either oriented along the $z$-direction or along the $-z$-direction, whereas $D_2$ posits that the same spin is oriented along the $x$-direction or along the $-x$-direction. More precisely: $D_1$ states that in some instantiations of the SP in question, there is an electron at the said location whose spin points along the $z$-direction, whereas in the rest of its instantiations, the spin points along the $-z-$-direction; and analogously so for $D_2$. Now, taking the two descriptions at face value \textit{prima facie} implies that there are instantiations of our SP, wherein an electron is present at the said location, and whose spin e.g. points along the $z$-direction \textit{and} also points along the $x$-direction. But, as we stated beforehand, to the best of our quantum-mechanically informed knowledge, there is no physically possible SP in which such an event can take place deterministically. We are thus caught up in a dilemma: (a) either we accept the taking place of seemingly absolutely indeterministic events, or (b) we take all the adequate descriptions ontologically seriously, but with the seemingly impossible caveat that the descriptions cannot be integrated in the way ordinary reasoning would suggest. The first horn - besides the relatively problematic bullets it needs to bite due to the aforementioned no-go theorems by Kochen, Specker and others - suffers from the positing of absolutely indeterministic events. 

The second horn however also suffers from a pathology, albeit a different one, in that it posits a seemingly unintelligible restriction. The quantum-logician might attempt to account for this restriction by saying that the logical law of distributivity of propositional conjunction and disjunction does not hold as generally as we might have thought.\footnote{The quantum-logical propositional system I am referring to here was first introduced by Birkhoff \& Von Neumann (1936) and was most influentially pushed as relevant for questions concerning the ontology of quantum phenomena by Putnam (1969).} This however does not seem (!) to be an account that could ever placate our worries, but only a redescription - or in this case, a formalization - of our worry. Let me note here that the rendition of the ``consistent histories'' formulation of quantum theory proposed by Robert Griffiths (1984) - a formulation that posits structures very similar to our PE-models above - faces the same conundrum: the various incompatible histories-narratives are all supposed to describe the same phenomenon, and yet there is a ``logical'' restriction on how these histories are to be combined. Whereas one is definitely free to come up with restrictions and rules that are to guide physicists in their practice, the ontologist is obliged to critically ask for a justification-explanation-elucidation of these restrictions.\footnote{One might urge that pushing for a reform of some seemingly widely implicitly or explicitly operative rule of reasoning cannot be more than a mere legislation of a convention regarding how certain words are to be used, and whose ``meanings'' differ from the usual `and's and `or's that appear in our ordinary (scientific) lives (something in the vicinity of this is nowadays advocated e.g. by Maudlin, 2005 and Warren, 2018). If that is right, any such push is to turn out either inconsistent - if the introduced conventions are intended to replace the ones implicitly operating in our ordinary practices - or empty - if the conventions are intended to merely lie parallelly to the ones already present, as merely instituting another manner of speaking. This conservativist critique should arguably extend also to attempts that modify the ``ordinary'' Kolmogorovian-like rules of probabilistic reasoning (see e.g. Feintzeig \& Fletcher, 2017) and perhaps even to suggestions that aim to salvage a form of Reichenbach's common-cause principle by positing ``quantum causes'' (Shrapnel, 2019). While I do think there is at least something to this critical view, an adequate reflection thereon would need to confront itself with the infinite depths of logic and of ``rules of reasoning''.}

Perhaps the message we ought to take from all of this is that the hermeneutic digestion we were pushing for - an understanding of quantum SPs in terms of ordinary SPs - reaches a limit that manifests itself in a seemingly incoherent situation. The project of finding an event-ontology of quantum phenomena is a coherent one, but its resolution is after all not as coherent as we, traditional ontologists, might have wanted. Quantum SPs just cannot be integrated more than this with ordinary SPs. It might be that our situation can be alleviated by perspectivalist-like proposals that I briefly mentioned above, the ones that tend to radically reform the ontological project that we are after; but this is not the right occasion to make any verdict on that. Let me remind you now that we have so far investigated merely one peculiar property of merely one subclass of quantum SPs. It will definitely be fruitful to continue the investigation, rather than to stop upon being ontologically stunned by its first obstruction: fruitful both for our understanding of the quantum, and, as mentioned at the beginning, for an occasional (!) help when it comes to its integration with relativity theory.

\subsection*{IV.III. Holistic Overtones}

In what follows we will briefly analyze a subclass of those quantum SPs to which multiple spatiotemporal trajectories $u_1,...,u_n:[0,1] \rightarrow \mathbb{R}^4$ can be associated and along which multiple relatively well localized non-relativistic quantum systems are posited to travel (think e.g. of an EPR-like experiment on electronic spins). One immediate mark that differentiates QM-models of such phenomena from those studied in the previous section is that quantum states assigned to composite systems can be ``entangled'' relative to the decomposition into the aforementioned spatially separated subsystems. While entanglement has on many occasions been taken to imply various kinds of relatively unexpected ontological consequences - e.g. various forms of ``holisms''\footnote{See Ismael \& Schaffer (2020) and references therein.} - we ought to recall that the distance between a QM-model of a SP and viable PE-models thereof is rather wide, so that peculiar features of the former may as well translate into non-peculiar features of the latter. Nevertheless, I will now indicate that there \textit{is} in fact a holistic component that is characteristic of PE-models of phenomena involving multiple systems, and that this is so \textit{not only} due to entanglement. 

Consider an experiment that features two devices. The first device tosses two fair coins at spatiotemporal location $x$, and depending on the outcomes $a\equiv (a_1,a_2)$, emits two electrons whose spin degrees of freedom are in one of the four mutually orthogonal Bell-states
\begin{align*}
    \ket{\phi^{(\pm)}}=\frac{1}{\sqrt{2}}\left(\ket{\uparrow}\ket{\uparrow}\pm\ket{\downarrow}\ket{\downarrow}\right)\\
    \ket{\psi^{(\pm)}}=\frac{1}{\sqrt{2}}\left(\ket{\uparrow}\ket{\downarrow}\pm\ket{\downarrow}\ket{\uparrow}\right)
\end{align*}
where states $\left\{\ket{\uparrow},\ket{\downarrow}\right\}$ correspond to the corresponding electron being oriented parallelly or antiparallelly to the $z$-direction. The two electrons are constrained to travel along trajectories $u_1,u_2:[0,1]\rightarrow \mathbb{R}^4$ - with the environments being so designed as to leave the electrons' state intact - until they encounter each other at spatiotemporal location $y$ and interact with the second device. The latter device implements a PVM capable of ideally perfectly distinguishing the 4 Bell states, thereby producing an outcome $b$ that is perfectly correlated to $a$, i.e. $P(a,b)=\frac{1}{4}\delta_{a,b}$.

Again, any PE-model $\mu$, if it is to be an adequate description of the above phenomenon, needs to be compatible with the distribution of the inputs $a$ and outputs $b$, i.e. $\mu_{\left\{x,y \right\}}(a,b)=\frac{1}{4}\delta_{a,b}$. The model may also trivially posit that at all locations along trajectories $u_1$ and $u_2$, events take place deterministically that correspond to the presence of an electron at that location. However, it is simple to notice that the model cannot posit any more informative events along the trajectories - such as an event that would correspond to the presence of an electron whose spin is oriented, say, parallelly to the $z$-direction - in that any such assignment would either be incompatible with the marginal distribution $P(a,b)$ or would violate Principle 2. Indeed, if the perfect correlation between $a$ and $b$ is to be preserved, then any random event posited in between them - e.g. at some location $u_1(t)$, for $t \in \langle 0,1\rangle$ - needs to correspond to a PVM that does not disturb the electrons' state, which is equivalent to saying that the PVM's elements need to commute with the projectors on the Bell states. Since there is no non-trivial projector decomposable as $\Pi \otimes \mathds{1}$ that commutes with any of the Bell states, then no random event of the aforementioned sort can be posited by an adequate PE-model. Now, the reason I am pointing out this feature is not due to its potential intrinsic interest, but due to the peculiarity of one of its consequences, which reads: according to any adequate PE-model of our phenomenon, there can be no collection of random events $\mathbf{c}$, posited to occur somewhere between locations $x$ and $y$, that could mediate the correlation between events $a$ and $b$ in the sense of $\mu(a,b,\mathbf{c})=\mu(b|\mathbf{c})\mu(\mathbf{c}|a)\mu(a)$. In other words, there is no adequate PE-model that portrays the causal influence between $a$ and its future random event $b$ as a product of a possibly continuous plurality of causal influences that obtain between them. We might tentatively denote this as the failure of \textit{causal continuity}, formalized as follows.

\bigskip

\textbf{Definition} (Causal continuity) A SP is \textit{causally continuous} if it admits of an adequate PE-model $\mu$ that satisfies the following property. For any $x,y \in \mathbb{R}^4$, such that $y \in \mathcal{C}_x^{(+)}$, there is a set $S \subset \left(\mathcal{C}_x^{(+)} \cap \mathcal{C}_y^{(-)}\right)$, for which it holds that
\begin{equation*}
    \mu_{\left\{x,y \right\} \cup S} (\omega_x,\omega_y,\omega_{S})=\mu_y(\omega_y|\omega_S)\mu_S(\omega_S|\omega_x)\mu_x(\omega_x),
\end{equation*}
for all $\omega_x \in \Omega_x,\omega_y \in \Omega_y,\omega_S \in \Omega_S$. A PE-model $\mu$ that satisfies the latter property is said to be causally continuous.

\bigskip

Note that the class of phenomena analyzed in the previous subsection - the ones that involve only relatively well localized single systems - all appear to be causally continuous. This is so because they admit of causally continuous PE-models that posit events that correspond to PVMs that commute with the momentary quantum states of the thematized systems; more precisely, at each location $u(t)$, an event is posited that corresponds to outcomes of a PVM whose elements all commute with $\rho(t)$. Causal continuity then follows from the continuity of the dynamical equations of closed quantum systems (the Schrödinger or Heisenberg equations): the quantum state at time $t$ is fully determined by the state at any $t'<t$. This also highlights the reason that our experiment with two systems prepared in Bell states fails to be causally continuous: the required PE-model would need to posit events that, for each $t$, correspond to outcomes of a PVM that consists of projectors on Bell states, and these outcomes cannot be associated individually to regions $u_1(t)$ and $u_2(t)$, but potentially only jointly to the composite region $\left\{ u_1(t),u_2(t)\right\}$. A form of causal continuity may thus be tentatively salvaged only by accepting the obtaining of \textit{non-separable} events, that is, events that neither occur in relatively well localized spatiotemporal regions, nor are they composed out of events that occur in such regions, but are instead understood to take place in an irreducible conjunction of two causally separated regions, regardless of the magnitude of their separation. It thus seems that a form of either spatial or temporal holism is inevitable: we are faced either with the acceptance of the physical possibility of the obtaining of non-separable events, or with the acceptance of the obtaining of causally related events that are not connected by a chained plurality of other causally connected events. 

The holistic aspect of PE-models of quantum SPs has above been exhibited on an example of paradigmatically entangled states. However, the same feature is also present in some experiments whose QM-models posit only separable quantum states. Take a variant of the experiment above, wherein two particles are constrained to travel along two trajectories, but instead of being prepared in one of the 4 Bell states, they are prepared equiprobably in one of $n$ separable orthogonal states $\left\{\rho_1,...,\rho_n\right\}$, before impinging on a device that distinguishes them perfectly. A necessary condition for this experiment to be causally continuous is for the $n$ states to be ``locally discriminable'', i.e. that the states be distinguishable on the basis of information gathered in an experiment in which the two particles are not allowed to physically interact: indeed, if this condition were not to hold, then there would in particular also be no local measurements on the particles whose outcomes could determine the outcome of the final joint measurement.\footnote{In quantum-information-theoretic language, ``experiments in which the two particles are not allowed to physically interact'' are said to consist only in ``local operations and classical communication'' (LOCC).} Interestingly, it has been shown that there is a plethora of separable and yet locally non-discriminable states: e.g. the first discoverers of this phenomenon (Bennett \textit{et al}., 1999) - who named it ``nonlocality without entanglement'' - proposed $9$ such states defined on two 3-dimensional systems (thus not realizable with electronic spins, but e.g. with photonic angular momenta). It therefore turns out that the causal discontinuity and its associated spatial-or-temporal holism that we beforehand spotted in experiments involving Bell states also extends to experiments that do not feature any entanglement between spatially separated systems: again, salient features of QM-models do not transpose one-to-one into features of PE-models, which is to say that the job of the ontologist is significantly harder than just reading off the representational correlate of some conveniently isolated gear pertaining to the quantum-mechanical formalism (`ontology' $\neq$ `what do mathematical objects $x,y,z$ represent?'). 

Above I suggested that a form of causal continuity can be restored by allowing for the positing of events that occur non-separably in 2 separated regions. A question that quite naturally suggests itself is: Can this always be done with maximally two regions, or are there phenomena in which non-separability across $n\geq 3$ regions needs to be invoked if one is to preserve causal continuity? What is the extent of the holism pertaining to (PE-models of) quantum SPs? There are some recent studies that indicate that such stronger non-separable phenomena are physically possible - generalizations of the above experiments with more particles travelling along more trajectories - in that a cognate of the prior necessary condition for causal continuity appears to fail: it seems that there are sets of mutually orthogonal separable states of $n$ particles that cannot be distinguished unless all of them are allowed to physically interact with each other (Halder et al., 2019; Zhou et al., 2023). However, an appropriately detailed analysis that carefully examines all of the assumptions involved in the argument is left for a future occasion. Let me now again cautiously remind you that we have so far only talked of holistic components of \textit{PE-models} of quantum SPs, and that it is far from clear what is the ontological lesson that is to be taken from this, i.e. how are we best to interpret our descriptive instrument now that it is once again displaying anomalous behavior. I will not delve into this here either, except for noting that the positing of occurrences of non-separable events does demand some serious ontological chewing, and perhaps (!) more so than its temporal counterpart, the failure of causal continuity, as this latter asserts something unexpected (causal discontinuity) about the ordinary (spatiotemporally localized events), instead of trying to retain the expected (causal continuity) by positing the unordinary (non-separable events). Again, the problematicity of the ``unordinary'' here is not its extravagance but its seeming lack of intelligibility. What is it for an event to occur non-separably across separated regions? What is the relationship between our ordinary events that can be spatiotemporally decomposed into pluralities of events, and these monolithic and yet spatiotemporally extended non-separable ``events''? This a genuine question, and not a merely rethorical one, one that a scientific ontology needs to address if it is to evaluate how it is best to cope with the here sketched holisms.

I want to conclude this chapter with yet another observation that will again not lead to the ample reflection that it deserves, and that will concern ``Bell's nonlocality'', i.e. the possibility of statistical correlations between distant events that cannot be traced back to a common cause without conflicting with some theoretical desideratum that we should arguably cling to (such as for instance the ``statistical independence'' of settings of apparently independently operating distant devices).\footnote{For a careful treatment of these theoretical desiderata, see Wiseman \& Cavalcanti (2017) and references therein.} In fact, it is hard to find nowadays a discussion on ontology and the quantum that does not sooner or later focalize on this phenomenon, so it does seem pertinent to spend at least a few breaths on it here. What Bell's work taught us is that any kind of positing of common causes in Bell-nonlocal phenomena is logico-mathematically destined to be theoretically problematic. But I want to emphasize that this positing has anyway always already been problematic, due to its ad hoc speculation on events whose occurrences are either absolutely indeterministic or are unsupported by current empirical evidence. Bell has thus revealed the structure of the ad-hoc-speculator's space of possibilities and provided more logico-mathematical cement that is to mark it as a no-go-zone. Positively, Bell-nonlocal phenomena, but also many EPR-like Bell-local phenomena, seem, according to their adequate PE-models, to exhibit the curious feature of lacking events that could serve as common causes of otherwise seemingly inexplicable correlations between distant events. This appears to be yet another curious datum concerning quantum SPs, along with the previously sketched complementarity and holism, that we need to not only learn how to ontologically cope with - inviting so to say for a vertical reflection - but to also better understand horizontally, e.g. by relating them to each other, explaining them in terms of each other, and by zooming into their fine-grained structure, as quantum physicists have indeed been doing for the last couple of decades.

\section*{V. Closing Remarks}

As announced at its beginning, this essay was an attempted contribution to our understanding of quantum phenomena by treating them as parallelly as possible to more ordinary and seemingly less problematic non-quantum phenomena, subsuming them all under one and the same broad category of SPs. The aim was to sit in a room, observe a coin toss in one of its corners, a quantum measurement in another, and portray both of these happenings on the same canvas, with the same paint and with the same technique. This integration of the unordinary and the ordinary was of course from its very start a regulative ideal that was impossible to reach. The desired parallellism was destined to be obstructed by anomalies that were to cause a divergence between the two paintings; after all, it would be unreasonable to expect the by now widely perceived paradoxical nature of the quantum to turn out being no more than a product of socio-historically contingent enthusiasm, and for its kernel to be best stored in the historian's rather than the physicist's closet. Still, despite its unattainability, the ontologist should keep reaching for this ideal and keep the said divergence at a healthy minimum.

In our attempted digestion of a handful of examples of quantum SPs we were accordingly pushed to accept certain peculiarities of their PE-models that seem to translate into event-ontological peculiarities: their irreducible complementarity, their causal discontinuity (or spatial non-separability) and the briefly mentioned absence of common causes of distant correlated random events. The contemporary reader has of course already encountered various renditions of all of these motives, which have indeed been appearing in conjunction with the quantum since its very beginnings. Still, I hope that this essay has managed to contribute to a more rigorous regimentation thereof, by arguing for the ways in which they are conditionally forced upon us in a reflection on event-occurrences in quantum SPs, and by formulating them by reference to \textit{prima facie} ontologically palatable events and their statistical correlations.

It is needless to say that the examples that have been surveyed here hardly exhaust even a corner of the vast plurality of phenomena that quantum theory informs us on - leaving out for instance the briefly mentioned categories taxonomized as phenomena involving non-localized quantum systems and quantum fields. This brings me to the promisory `future' alluded to in the title and to the afore-made promisory note concerning a potential help when it comes to an integration of quantum and relativity theory. The point of ontology, qua systematic science, is not to deal with one particular here and another observation there, but to erect a systematic whole. One of the goals of a quantum event-ontological investigation is the crystallization of a set of principles that, albeit suggested by quantum phenomena and by the grip we have on them in terms of QM-models, appear to be valid universally and can respectively be phrased in terms that do not have anything in particular to do with our current formulations of quantum theory, and that would accordingly be codifiable mathematically in terms of properties of general PE-models, the same kind of models that can be used to describe ordinary non-quantum phenomena. I do not know whether such principles, that for now lie dormant in an ideal future, would refer among other things to the here thematized possibility of irreducible complementarities and causal discontinuities. It is premature to speculate on this in that the actually systematic work is still to be done - here only examples and generalized observations have been given. Moreover, if such an investigation is to have any chance of being fruitful when it comes to connections with relativity, the treatment of space and time will certainly need to overcome the naivety of their treatment in this essay, and will need to take into account relativizations to reference frames, coordinate systems, general-relativistic modifications, and so forth.

A search for general principles that are hinted at by quantum theory but that can be phrased in more ordinary non-quantum-mechanical terms is reminiscent of the projects that seek to derive certain structural features of the quantum-mechanical formalism from constraints imposed on ``general probabilistic theories''.\footnote{See Müller (2021), Plávala (2023) and references therein.} The aims of these projects however diverge from the ones pointed at by this essay in that general probabilistic theories are identified by reference to systems, measurements and other operations, whose home is not ontology but information theory. Physical ontology, that is, has neither place for semiotic distinctions between measurements and non-measurements, operations and non-operations, nor for epistemological distinctions between theoretical and observational vocabulary or between unobservable and observable events. While the `it' has a lot to learn from the `bit', the study of the former is not to be replaced, but only complemented by the study of the latter. And - recalling what was said at the beginning - this reaching for the `it', for the physically real, is by no means a dogma or a contingently suggested modern-scientific thesis, an \textit{ism} that we might as well learn to ignore and finally supersede. It is rather an expression of a deep hermeneutic need of the physicist, of the basic spirit that has been animating the project of physics \textit{qua} science of this very world in which we ourselves are embodied. This spirit can however not be satisfied by opaque proclamations of the representational correctness of highly mathematized theories without adequately elucidating the relation between their physical correlate and the `it' that we encounter in our everydayness. What we thus need, in summary, is a synthesis of the Platonic and of the Nietzschean drive - yes, a reaching for what really is the case, for the truth; but a truth that speaks not only to the Gods but also to us, human beings.

\newpage

\section*{Funding Information}
The author acknowledges the Austrian Science Fund (FWF) for its financial support through Grants 10.55776/DOC162 and 10.55776/P36994.

\section*{References}

Adlam, E., \& Rovelli, C. (2023). Information is Physical: Cross-Perspective Links in Relational Quantum Mechanics. \textit{Philosophy of Physics, 1}(1).

\smallskip

Adlam, E. (2024). Moderate Physical Perspectivalism. Philosophy of Science, 1-21.

\smallskip

Bennett, C. H., DiVincenzo, D. P., Fuchs, C. A., Mor, T., Rains, E., Shor, P. W., ... \& Wootters, W. K. (1999). Quantum nonlocality without entanglement. \textit{Physical Review A, 59}(2), 1070.

\smallskip

Birkhoff, G., \& Von Neumann, J. (1936). The Logic of Quantum Mechanics. \textit{Annals of Mathematics, 37}(4), 823–843.

\smallskip

Di Biagio, A., \& Rovelli, C. (2022). Relational quantum mechanics is about facts, not states: A reply to Pienaar and Brukner. \textit{Foundations of Physics, 52}(3), 62.

\smallskip

Dieks, D. (2022). Perspectival quantum realism. \textit{Foundations of Physics, 52}(4), 95.

\smallskip

Feintzeig, B. H., \& Fletcher, S. C. (2017). On noncontextual, non-Kolmogorovian hidden variable theories. \textit{Foundations of Physics, 47}, 294-315.

\smallskip

Fuchs, C. A. (2023). QBism, Where Next?. In \textit{Phenomenology and QBism} (pp. 78-143). Routledge.

\smallskip

Griffiths, R. B. (1984). Consistent histories and the interpretation of quantum mechanics. \textit{Journal of Statistical Physics, 36}, 219-272.

\smallskip

Halder, S., Banik, M., Agrawal, S., \& Bandyopadhyay, S. (2019). Strong quantum nonlocality without entanglement. \textit{Physical review letters, 122}(4), 040403.

\smallskip

Healey, R. (2017). The quantum revolution in philosophy. \textit{Oxford University Press}.

\smallskip

Ismael, J., \& Schaffer, J. (2020). Quantum holism: Nonseparability as common ground. \textit{Synthese}, 197, 4131-4160.

\smallskip

Maudlin, T. (2005). The tale of quantum logic. \textit{Hilary Putnam}, 156-187.

\smallskip

Lewis, P. J. (2016). \textit{Quantum ontology: A guide to the metaphysics of quantum mechanics}. Oxford University Press.

\smallskip

Lombardi, O. I., \& Dieks, D. (2017). Modal interpretations of quantum mechanics. \textit{The Stanford Encyclopedia of Philosophy}.

\smallskip

Massimi, M. (2022). \textit{Perspectival Realism}. Oxford University Press

\smallskip

Müller, M. (2021). Probabilistic theories and reconstructions of quantum theory. \textit{SciPost Physics Lecture Notes}, 028.

\smallskip

Nguyen, J., \& Frigg, R. (2022). \textit{Scientific representation}. Cambridge University Press.

\smallskip

Plávala, M. (2023). General probabilistic theories: An introduction. \textit{Physics Reports, 1033}, 1-64.

\smallskip

Putnam, H. (1969). Is logic empirical?. In \textit{Boston Studies in the Philosophy of Science: Proceedings of the Boston Colloquium for the Philosophy of Science 1966/1968} (pp. 216-241). Dordrecht: Springer Netherlands.

\smallskip

Shrapnel, S. (2019). Discovering quantum causal models. \textit{The British Journal for the Philosophy of Science}.

\smallskip

Spekkens, R. W. (2007). Evidence for the epistemic view of quantum states: A toy theory. \textit{Physical Review A—Atomic, Molecular, and Optical Physics, 75}(3), 032110.

\smallskip

Tao, T. (2011). \textit{An introduction to measure theory} (Vol. 126). American Mathematical Society.

\smallskip

Valentini, A. (2023). Beyond the Born rule in quantum gravity. \textit{Foundations of Physics, 53}(1), 6.

\smallskip

Van Fraassen, B. C. (2008). \textit{Scientific representation: Paradoxes of perspective}. Oxford University Press.

\smallskip

Warren, J. (2018). Change of logic, change of meaning. \textit{Philosophy and Phenomenological Research, 96}(2), 421-442.

\smallskip

Werner, R. F. (1989). Quantum states with Einstein-Podolsky-Rosen correlations admitting a hidden-variable model. \textit{Physical Review A, 40}(8), 4277.

\smallskip

Wiseman, H. M., \& Cavalcanti, E. G. (2017). Causarum Investigatio and the two Bell’s theorems of John Bell. \textit{Quantum [Un] Speakables II: Half a Century of Bell's Theorem}, 119-142.

\smallskip

Zhou, H., Gao, T., \& Yan, F. (2023). Strong quantum nonlocality without entanglement in an n-partite system with even n. \textit{Physical Review A, 107}(4), 042214.

\end{document}